\documentclass[manuscript]{aastex}

\usepackage{epstopdf}

\shorttitle{Semi-Major Axis Drifts of Near-Earth Asteroids}
\shortauthors{Nugent and Margot}

\begin{document}


\title{Detection of Semi-Major Axis Drifts in 54 Near-Earth Asteroids:
       New Measurements of the Yarkovsky Effect}


\author{C. R. Nugent\altaffilmark{1}, J. L. Margot\altaffilmark{1,2}, S. R. Chesley,\altaffilmark{3} and D. Vokrouhlick{\'y}\altaffilmark{4} }
\altaffiltext{1}{Department of Earth and Space Sciences, University of California, Los Angeles, CA 90095, USA}
\altaffiltext{2}{Department of Physics and Astronomy, University of California, Los Angeles, CA 90095, USA}
\altaffiltext{3}{Jet Propulsion Laboratory, California Institute of Technology, Pasadena, CA 91109, USA}
\altaffiltext{4}{Institute of Astronomy, Charles University, V Hole\u{s}ovi\u{c}k\'{a}ch 2, CZ-18000 Prague 8, Czech Republic}

\begin{abstract}
We have identified and quantified semi-major axis drifts in Near-Earth
Asteroids (NEAs) by performing orbital fits to optical and radar
astrometry of all numbered NEAs.  We focus on a subset of 54 NEAs that
exhibit some of the most reliable and strongest drift rates.  Our
selection criteria include a {\em Yarkovsky sensitivity} metric that
quantifies the detectability of semi-major axis drift in any given
data set, a signal-to-noise metric, and orbital coverage requirements.
In 42 cases, the observed drifts ($\sim 10^{-3}$ AU/Myr) agree well
with numerical estimates of Yarkovsky drifts.  This agreement suggests
that the Yarkovsky effect is the dominant non-gravitational process
affecting these orbits, and allows us to derive constraints on
asteroid physical properties.  In 12 cases, the drifts exceed nominal
Yarkovsky predictions, which could be due to inaccuracies in our
knowledge of physical properties, faulty astrometry, or modeling
errors.  If these high rates cannot be ruled out by further
observations or improvements in modeling, they would be indicative of
the presence of an additional non-gravitational force, such as that
resulting from a loss of mass of order a kilogram per second.  We
define the {\em Yarkovsky efficiency} $f_Y$ as the ratio of the change
in orbital energy to incident solar radiation energy, and we find that
typical Yarkovsky efficiencies are $\sim$10$^{-5}$.
\end{abstract}

\keywords{astrometry --- minor planets, asteroids --- minor planets,
  asteroids: individual (1999 RQ36, Aten, Apollo, Ganymed, Geographos,
  Hathor, Icarus, Orpheus, Ra-Shalom) --- radiation mechanisms:
  thermal}

\section{Introduction}

Understanding how the Yarkovsky force modifies asteroid orbits has
illuminated how asteroids and meteorites are transported to near-Earth
space from the main belt and has allowed for deeper understanding of
the structure of asteroid families~\citep{2006AREPS}. The Yarkovsky
force is necessary for accurately predicting asteroid trajectories,
including those of potentially hazardous asteroids
\citep{gior02,ChesleyHazard,gior08,MilaniImpact1999RQ36}.

The Yarkovsky effect (or force) describes the process by which an
asteroid's surface thermal lag and rotation result in net thermal
emission that is not aligned towards the Sun
\citep{BottkeYarkAstIII,2006AREPS}.  The so-called {\em diurnal
  component} of the Yarkovsky effect operates as follows.  A
prograde-spinning object generally has a component of this surface
thermal emission anti-aligned with the motion along the orbit,
producing a net increase in the object's semi-major axis (i.e., $da/dt
> 0$, where $a$ is the semi-major axis).  Conversely, a
retrograde-spinning object generally has a component aligned with its
velocity, shortening its semi-major axis (i.e., $da/dt < 0$).

The maximum possible drift rate for any radiation-powered force acting
on near-Earth asteroids (NEAs) can be obtained by equating the
incident solar radiation energy in a given time interval to the change
in orbital energy during the same interval.  We find
\begin{equation}
\label{equ:dadtmax}
\frac{da}{dt} = f_Y \frac{3}{4\pi}\frac{1}{\sqrt{1-e^2}}\frac{L_\odot}{GM_\odot}\frac{1}{D\rho},
\end{equation}
where $f_Y$ is an efficiency factor analogous to that used by
\citet{GoldreichSari}, $e$ is the eccentricity, $L_\odot$ and
$M_\odot$ are the luminosity and mass of the Sun, $G$ is the
gravitational constant, and $D$ and $\rho$ are the effective diameter
and bulk density of the asteroid.  This equation exhibits the expected
dependence on the asteroid area-to-mass ratio.  In convenient units,
it reads
\begin{equation}
\label{equ:dadttyp}
\frac{da}{dt} = \frac{1.457}{\sqrt{1-e^2}}\left(\frac{f_Y}{10^{-5}}\right)\left(\frac{1\ {\rm km}}{D}\right)\left(\frac{1000\ {\rm kg \ m}^{-3}}{\rho}\right) 10^{-3} {\rm AU/Myr}.
\end{equation}
Maximum efficiency ($f_Y$=1) would convert all incoming solar
radiation into a change in orbital energy.  We will show in
Section~\ref{sec:results} that typical Yarkovsky efficiencies are $f_Y
\sim 10^{-5}$, and that typical rates are $\sim 10^{-3} $AU/Myr for
kilometer-sized asteroids.  The low efficiency and rates are due to
the fact that it is the momentum of departing thermal photons that
moves the asteroid.

\citet{ChesleyGolevka} used precise radar ranging measurements to
(6489) Golevka and reported the first detection of asteroidal
Yarkovsky drift.  The drift rate for this NEA of $da/dt = (-6.39 \pm
0.44) \times 10^{-4}$ AU/Myr~\citep{ChesleyLCMfits} corresponds to an
efficiency $f_Y = 5\times 10^{-6}$ for $D$=530~m and
$\rho=2700$~kg~m$^{-3}$.

\citet{Vok1992BF} employed the Yarkovsky effect to link a 1950
observation to asteroid (152563) 1992 BF with a $da/dt$ rate of
$(-10.7 \pm 0.7) \times 10^{-4}$ AU/Myr.  This corresponds to an
efficiency $f_Y = 7\times 10^{-6}$ for $D$=420~m and
$\rho$=2500~kg~m$^{-3}$.
If 1992 BF has a density closer to 1500~kg~m$^{-3}$, the efficiency
would be $f_Y = 4\times 10^{-6}$.

There have been other searches for the effects of non-gravitational
forces in asteroid orbits.  \citet{sitarskiIcarus} considered a
semi-major axis drift in the orbit of (1566) Icarus and found $da/dt =
(-7.3 \pm 3.9) \times 10^{-4}$ AU/Myr.  Our best estimate is $da/dt=
(-3.2 \pm 2.0) \times 10^{-4}$ AU/Myr.  \citet{sitarskiToutatis} found it
necessary to incorporate a non-gravitational term $da/dt = -58 \times
10^{-4}$ AU/Myr in his orbit determination of (4179) Toutatis,
however the availability of radar ranges in 1992, 1996, 2004, and 2008
strongly suggest a drift magnitude that does not exceed $-5 \times
10^{-4}$ AU/Myr.  \citet{Ziol} examined the orbits of 10 asteroids and
found drifts in four asteroids, including a $(-295.7 \pm 14.6) \times
10^{-4}$ AU/Myr drift for (1862) Apollo.
\citet{Yeomanscomet} used a cometary model to search for perturbations
and also detected a drift associated with (1862) Apollo, though a
value was not reported. Our best estimate is $(-2.38 \pm 0.25) \times
10^{-4}$ AU/Myr (Section \ref{sec:results}).  It appears that these
early estimates are not aligned with modern determinations, and may
have been caused by erroneous or insufficient astrometry.  More
recently, \citet{ChesleyLCMfits} searched for Yarkovsky signatures and
reported rate estimates for 12 candidates.

Here we use new developments in star catalog
debiasing~\citep{ChesleyBias} as well as the most recent astrometric
data to compute semi-major drift rates for select NEAs, which
multiplies the number of existing measurements by a factor of $\sim$4.

Observations of Yarkovsky rates can be used to place constraints on
composition (i.e.\ metal vs.\ rock), physical properties (i.e.\ bulk
density), and spin properties (i.e.\ prograde vs.\ retrograde).  The
magnitude of the force is dependent on the object's mass, size,
obliquity, spin rate, and surface thermal properties. Separating how
each of these quantities uniquely contributes to a measured $da/dt$ is
often not possible, but past Yarkovsky detections have allowed for
insight into the associated objects. With certain assumptions on
surface thermal properties, bulk densities were determined from the
measured drifts of Golevka \citep{ChesleyGolevka} and (152563) 1992 BF
\citep{Vok1992BF}.  For the latter, the magnitude and direction of the
drift point to an obliquity in excess of 120
degrees~\citep{Vok1992BF}.  

\section{Methods}
\label{sec:methods}

\subsection{Yarkovsky sensitivity}
The Yarkovsky drift manifests itself primarily as a change in mean
anomaly (or along-track position), and some observational
circumstances are poorly suited to detect such changes.  Examples
include optical astrometry secured when the line-of-sight is roughly
parallel to the asteroid velocity vector or when the object is at
large distances from Earth.  In both instances the differences in
astrometric positions can be much smaller than observational
uncertainties, resulting in low sensitivity to the Yarkovsky effect.
The overall Yarkovsky sensitivity depends on the orbital geometry of
the NEA and on the entire set of available observations.  This can be
quantified rigorously.  For each epoch $t_i$ at which optical
observations were obtained ($1 \leq i \leq N$), we predict the
position $P_i^0$ for the best-fit orbit ($da/dt=0$) and the position
$P_i^*$ for the same orbit modified by a nominal non-zero $da/dt$.
The value of the nominal rate is not important as long as it results
in detectable ($\sim$arcsecond) changes in coordinates and as long as
it is applied consistently to all objects; we used $da/dt$=0.1 AU/Myr.

We then define the {\em Yarkovsky sensitivity} $s_Y$ as
\begin{equation}
s_Y = \sqrt{\frac{1}{N}\sum_{i=1}^N \frac{(P_i^*-P_i^0)^2}{\sigma_i^2}},
\label{eq-sy}
\end{equation}
where $\sigma_i$ is the positional uncertainty associated with
observation $i$.  This root mean square quantity provides an excellent
metric to assess the relative sensitivity of any given data set to a
drift in semi-major axis, including drifts caused by Yarkovsky
influences.  The metric can be applied to the entire set of available
observations, or to the subset of observations that survive the
outlier rejection steps described below.  We computed both quantities
and used the latter for our analysis.  We found that data sets with
scores $s_Y$ below unity yield unreliable results, including
artificially large rates and large error bars.  Out of $\sim$1,250
numbered NEAs, only $\sim$300 have $s_Y > 1$ and $\sim$150 have $s_Y >
2$.  In this paper we focus on a subset of these NEAs.

\subsection{Orbital fits}
For this work we employed orbital fits to optical astrometry to
determine semi-major axis drift rates for NEAs.  We used the OrbFit
software package, which is developed and maintained by the OrbFit
Consortium \citep{MilaniOrbitBook}.  OrbFit can fit NEA trajectories
to astrometric data by minimizing the root mean square of the weighted
residuals to the data, optionally taking into account a given non-zero
rate of change in semi-major axis $da/dt$.  We included perturbations
from 21 asteroids whose masses were estimated by \citet{kono11}.

We downloaded optical astrometry for all numbered minor planets
(NumObs.txt.gz) from the Minor Planet Center (MPC) on January 31st,
2012.  We have assumed that all the astrometry has been properly
converted to the J2000 system.  The quality of the astrometry varies
greatly, and we applied the data weighting and debiasing techniques
implemented in OrbFit, which appear to follow the recommendations of
\citet{ChesleyBias}.  Data weights are based on the time the
observation was performed, the method of the observation (CCD or
plate), the accuracy of the star catalog, and in some cases the
accuracy of the observatory.  Correction for known star catalog biases
was applied when possible.  Biases vary depending on the specific star
catalog and region of the sky, and can reach 1.5 arcseconds in both
right ascension (R.A.) and declination (Dec.).  Correction for these
biases can substantially improve the recovery of orbital parameters
from observations.  However, as discussed in \citet{ChesleyBias}, not
every observation can be debiased. Some observations were reported to
the MPC without noting the star catalog used in the data
reduction. Although \citet{ChesleyBias} deduced the star catalogs used
by several major surveys, there remain observations from smaller
observatories that do not have associated star catalogs.  Accordingly,
a fraction of the astrometry used in this paper was not debiased.
Based on counts published \citet{ChesleyBias}, we estimate this
fraction to be less than 7.2\% of all the observations.

Our procedure for determining the semi-major axis drift rate included
three steps: an initial fit to the debiased data, an outlier rejection
step, and a search for the best-fit $da/dt$, with iteration of the
last two steps when necessary.

We used the orbital elements from the Minor Planet Center's MPCORB
database as initial conditions for the first fit for each object (step
1). This first fit, performed with $da/dt = 0$ and outlier rejection
turned off, slightly corrected the orbital elements for our weighted,
debiased observations. The orbital elements from each object's first
fit became the starting orbital elements for all later fits of that
object.

The second fit of each object served to reject outliers and was
initially performed with $da/dt = 0$ (step 2). The residual for each
observation was calculated using the usual observed (O) minus computed
(C) quantities:
\begin{equation}
\chi_{\rm res}=\sqrt{\left(\frac{({\rm R.A.}_{\rm O}-{\rm R.A.}_{\rm C}) \times \rm cos(\rm Dec._{\rm O})}{\sigma_{{\rm R.A.}} }\right)^2+\left(\frac{{\rm Dec.}_{\rm O}-{\rm Dec.}_{\rm C}}{\sigma_{{\rm Dec.}}}\right)^2},
\end{equation}
where $\sigma_{{\rm R.A.}}$ and $\sigma_{{\rm Dec.}}$ are the
uncertainties for that observation in R.A.\ and Dec., respectively.
We rejected observations when their $\chi_{\rm res} >\sqrt{8}$, and
recovered previously rejected observations at $\chi_{\rm
res}=\sqrt{7}$, with the rejection step iterated to convergence.
Results are fairly robust over a large range of thresholds for
rejection (Section~\ref{sec:results}).  If the post-fit residuals were
normally distributed, the chosen thresholds would result in $< 1\%$ of
observations being rejected as outliers.  Because errors are not
normally distributed, our typical rejection rates are 2-5\% of all
available astrometry.  This second fit produced the set of
observations which were used in the third step.

The third step was a series of orbital element fits to the
observations over a set of fixed $da/dt$ values.  During these fits,
we used the set of observations defined by the second fit and did not
allow further outlier rejection.  The quality of a fit was determined
by summing the squares of residuals $\chi^2 = \sum \chi_{\rm res}^2$.
To locate the region with the lowest $\chi^2$, we used a three-point
parabolic fit or the golden-section minimization
routine~\citep{NumericalRec}.  A parabola was then fit to the $\chi^2$
curve in the vicinity of the minimum, and we used the minimum of the
parabola to identify the best-fit $da/dt$ value.

\begin{figure}[ht]
	\begin{center}
		\includegraphics[scale=1.0]{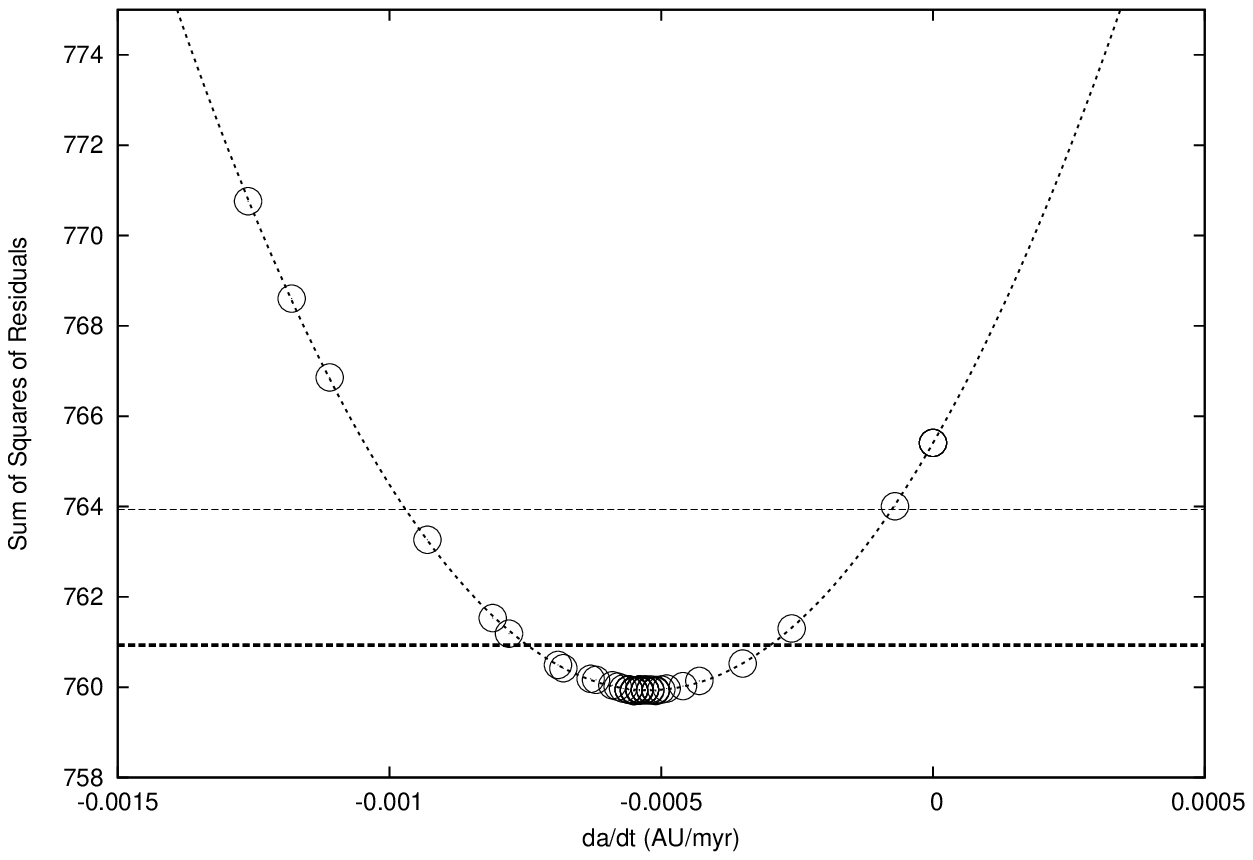}
	\end{center}
	\caption{Search for best-fit $da/dt$ value to optical
          astrometry of (2100) Ra-Shalom (1,281 observation epochs,
          2,562 observations, 7 adjustable parameters, 2,555 degrees
          of freedom).  The sums of squares of residuals corresponding
          to a range of $da/dt$ values are shown as circles, with a
          parabolic fit shown as a dotted line.  The $da/dt$ values
          plotted here were determined by the golden section search
          algorithm \citep{NumericalRec} as it searched for and found
          a minimum at $da/dt=-5.20 \times 10^{-4}$ AU/Myr with a
          reduced $\chi^2$ value of $0.30$.  Confidence limits of
          $68.3\%$ ($1\sigma$) are indicated by the thick dashed line,
          and correspond to the range $da/dt=[-7.4,-2.9] \times
          10^{-4}$ AU/Myr.  The thin dashed line shows the 95.4\%
          ($2\sigma$) confidence region.  
	\label{fig:2100chi}
}
\end{figure}

Confidence limits were estimated using $\chi^2$ statistics.
Confidence regions of 68.3\% and 95.4\% ($1\sigma$ and $2\sigma$,
respectively) were established by the range of $da/dt$ values that
yielded $\chi^2$ values within 1.0 and 4.0 of the best-fit $\chi^2$
value, respectively (Fig.~\ref{fig:2100chi}).

The initial outlier rejection step can in some cases eliminate valid
observations simply because the Yarkovsky influences are not captured
in a dynamical model with $da/dt=0$.  To circumvent this difficulty,
we iterated the outlier rejection step with the best-fit $da/dt$ value
and we repeated the fitting process.  In 52 out of 54 cases, the new
best-fit value matched the previous best-fit value to within
$1\sigma$, and we accepted the new best-fit values as final.  For the
other objects we repeated the reject and fit processes until
successive best-fit values converged within $1\sigma$ (which never
required more than one additional iteration).  Our results report the
$da/dt$ values obtained at the end of this iterative process.

\subsection{Sample selection}
\label{sec-select}

We restricted our study to numbered NEAs with the best Yarkovsky
sensitivity (Equation~\ref{eq-sy}), specifically $s_Y > 2$
(Fig.~\ref{fig-xmas}).

We also chose to focus on objects with non-zero $da/dt$ values by
using a signal-to-noise ratio (SNR) metric, defined as the ratio of
the best-fit $da/dt$ to its $1\sigma$ uncertainty.  We accepted all
objects with SNR $>1$ (Fig.~\ref{fig-xmas}).

Some asteroids have observations that precede the majority of the
object's astrometry by several decades and have relatively high
uncertainties. In order to test the robustness of our results, we
removed these sparse observations, which were defined as ten or fewer
observations over a 10-year period.  Fits were then repeated for
these objects without the early observations. If the initial best-fit
value fell within the $1\sigma$ error bars of the new best-fit
value, the initial result was accepted, otherwise, the object was
rejected.  

Superior detections of the Yarkovsky effect are likely favored with
longer observational arcs, larger number of observations, and good
orbital coverage.  For this reason we limited the sample to those NEAs
with an observational arc at least 15 years long, with a number of
reported observations exceeding 100, and with at least 8 observations
per orbit on at least 5 separate orbits.

We report on the 54 objects that met all of these criteria:
sensitivity, SNR, sparse test, and orbital coverage.

\subsection{Validation}
We validated our optical-only technique whenever radar ranging
observations were available on at least two apparitions.  This could
only be done for a fraction of the objects in our sample.  In the
remainder of this paper, {\em optical-only} results are clearly
distinguished from {\em radar+optical} results.  For the radar+optical
fits, we included all available radar astrometry and disallowed
rejection of potential radar outliers.  The internal consistency of
radar astrometry is so high that outliers are normally detected before
measurements are reported.

We also verified that a fitting procedure that holds successive
$da/dt$ values constant is equivalent to performing 7-parameter fits
(6 orbital parameters and $da/dt$ simultaneously).  The $da/dt$ values
obtained with both procedures are consistent with one another.

\subsection{Yarkovsky modeling}
In addition to the measurements described above, we produced numerical
estimates of the diurnal Yarkovsky drift for each of the objects in
our sample.  Comparing the measured and estimated rates provides a way
to test Yarkovsky models.  In some instances, e.g.,\ robust
observations irreconcilable with accurate Yarkovsky modeling, it could
also lead to the detection of other non-gravitational forces, such as
cometary activity.  Our numerical estimates were generated as follows.
At each timestep, we computed the diurnal Yarkovsky acceleration
according to equation (1) of \citet{VokSmallNEAS}, which assumes a
spherical body, with the physical parameters~\citep{OpeilTherCon}
listed in Table \ref{tab-thercon} and an assumption of 0$^\circ$ or
180$^\circ$ obliquity.  We assumed that the thermal conductivity did
not have a temperature dependence, but found that adding a
temperature-dependent term according to the prescription of
\citet{HutterRad} ($K=K_0 + K_1 T^3$, with $K_1 = 0.0076$) did not
change our predictions by more than 1\%.
We then resolved the acceleration along orthogonal directions,
and used Gauss' form of Lagrange's planetary equations \citep{Danby}
to evaluate an orbit-averaged $da/dt$.
  
The physical parameters chosen for these predictions mimic two
extremes of rocky asteroids; one is intended to simulate a rubble pile
with low bulk density, the other a regolith-free chunk of rock (Table
\ref{tab-thercon}).  These parameters correspond to a thermal inertia
range of $77 - 707$ J~m$^{-2}$~s$^{-0.5}$~K$^{-1}$, enveloping the
results of \citet{Delbo07ThermInertYark}, who found an average NEA
thermal inertia to be $200$~J~m$^{-2}$~s$^{-0.5}$~K$^{-1}$.  In most
cases, the drift rates produced by these two extreme cases encompass
the drift produced by a rubble-pile object that has a regolith-free
surface, or the drift produced by a solid object with regolith.

\begin{table}[h]
\begin{center}
	\caption{Physical and thermal properties used for numerical
          estimates of the semi-major axis drift of asteroids. Thermal
          properties are based on measurements of three meteorites at
          200~K, as measured by \citet{OpeilTherCon}. Listed are heat
          capacity $C$, thermal conductivity $K$, bulk density of the
          surface $\rho_s$, and mean bulk density
          $\rho$. \label{tab-thercon} }
\vspace{0.2cm}
\begin{tabular}{lcccc}
\tableline\tableline
Composition & $C$ (${\rm J\ kg}^{-1}{\rm K}^{-1}$)& $K$ (${\rm W\ m}^{-1} {\rm K}^{-1}$) & $\rho_s$ (${\rm kg\  m}^{-3}$ ) & $\rho$ (${\rm kg\  m}^{-3}$ ) \\ 
 \tableline
Rubble Pile & 500 & 0.01 & 1200 & 1200 \\ 
Rock Chunk  & 500 & 0.50 & 2000 & 2000\\ 
\tableline
\end{tabular}
\end{center}
\end{table}

There is no simple relationship between these physical parameters
and predicted drift rates, but for most cases the rubble pile exhibits
the larger $da/dt$ values due to its low bulk density
(Equation~\ref{equ:dadttyp}).  The smaller values of density of the
surface and thermal conductivity for rubble piles produce a smaller
thermal inertia, and therefore a longer thermal lag.
Generally, but not always, this longer thermal lag, combined with the
rotation of the asteroid, allows for a larger fraction of departing
thermal emission to be aligned with the asteroid's velocity, resulting
in a larger drift.

When available, measured values of the geometric albedo, diameter, and
spin rate from the JPL Small-Body Database \citep{SmallBodyDatabase}
were incorporated into our predictions for Yarkovsky drifts.  When not
available, the diameter $D$ in km was estimated from the absolute
magnitude $H$ using \citep{PravecHarrisIcarus},
\begin{equation}
\label{equ:diameter}
 D=\frac{1329}{\sqrt{p_{V}}} * 10^{-0.2H}
\end{equation} 
where we used two values of the V-band geometric albedo $p_V$ (0.05
and 0.45), a range that captures observed albedos for the majority of
NEAs.  When spin rate was unknown, we assumed a value of 5
revolutions/day, based on the average spin rate values for asteroids 1
to 10 km in diameter shown in Fig. 1 of
\citet{PravecFastSlow}. Emissivity was assumed to be 0.9.  Bond albedo
was estimated with a uniform value of the phase integral (q=0.39) on
the basis of the IAU two-parameter magnitude system for asteroids
\citet{bowe89} and an assumed slope parameter G=0.15.

We have assumed $p_V = 0.14$ for the purpose of quantifying the
Yarkovsky efficiency when the asteroid size was unknown.

\section{Results}
\label{sec:results}

\begin{figure}[ht]
	\begin{center}
		\includegraphics[width=5in]{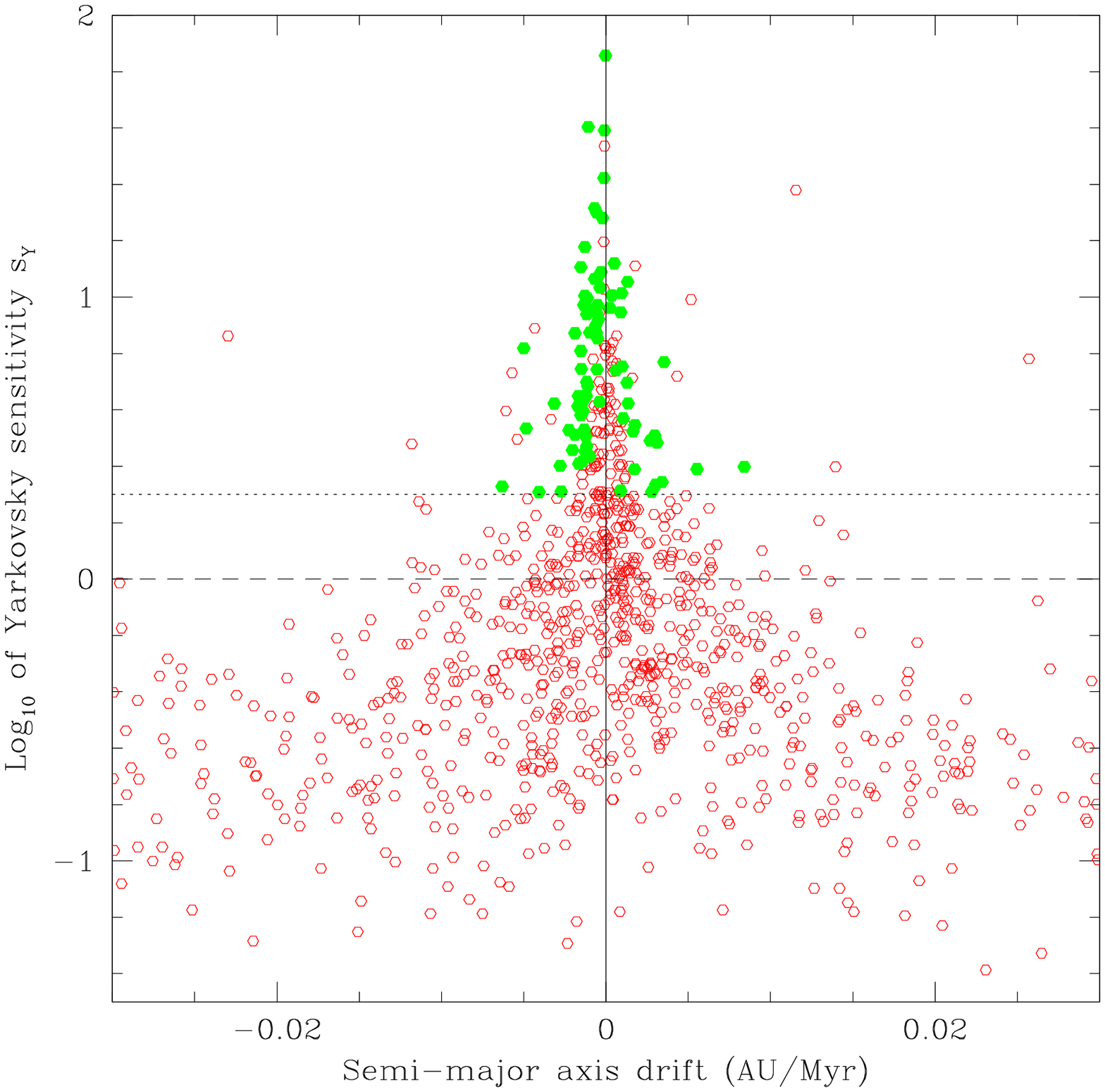}
	\end{center}
	\caption{Yarkovsky sensitivity metric $s_Y$ plotted as a
          function of semi-major axis drift rate $da/dt$ for 1,252
          numbered NEAs.  Data sets with Yarkovsky sensitivity below
          unity (dashed line) yield unreliable results, including
          large rates and large error bars.  Our selection criteria
          require $s_Y > 2$ (dotted line) and SNR $> 1$.  The 80
          objects that meet both selection criteria are shown in
          green.  About 26 of these 80 NEAs are eliminated by the
          sparse test and orbital coverage requirements (see
          Section~\ref{sec-select}).}
	\label{fig-xmas}
\end{figure}

We measured the semi-major axis drift rate of all 1,252 numbered NEAs
known as of March 2012.  Some of the drift rates are not reliable
because of poor sensitivity to Yarkovsky influences
(Fig.~\ref{fig-xmas}). 

After our process of selection and elimination
(Section~\ref{sec-select}), we were left with 54 NEAs that exhibit
some of the most reliable and strongest drift rates.  Although we
report objects with $s_Y > 2$, we have the most confidence in objects
with highest Yarkovsky sensitivity, and we show objects in order of
decreasing $s_Y$ value in our figures.

\begin{figure}[ht]
	\begin{center}
		\includegraphics[width=6in]{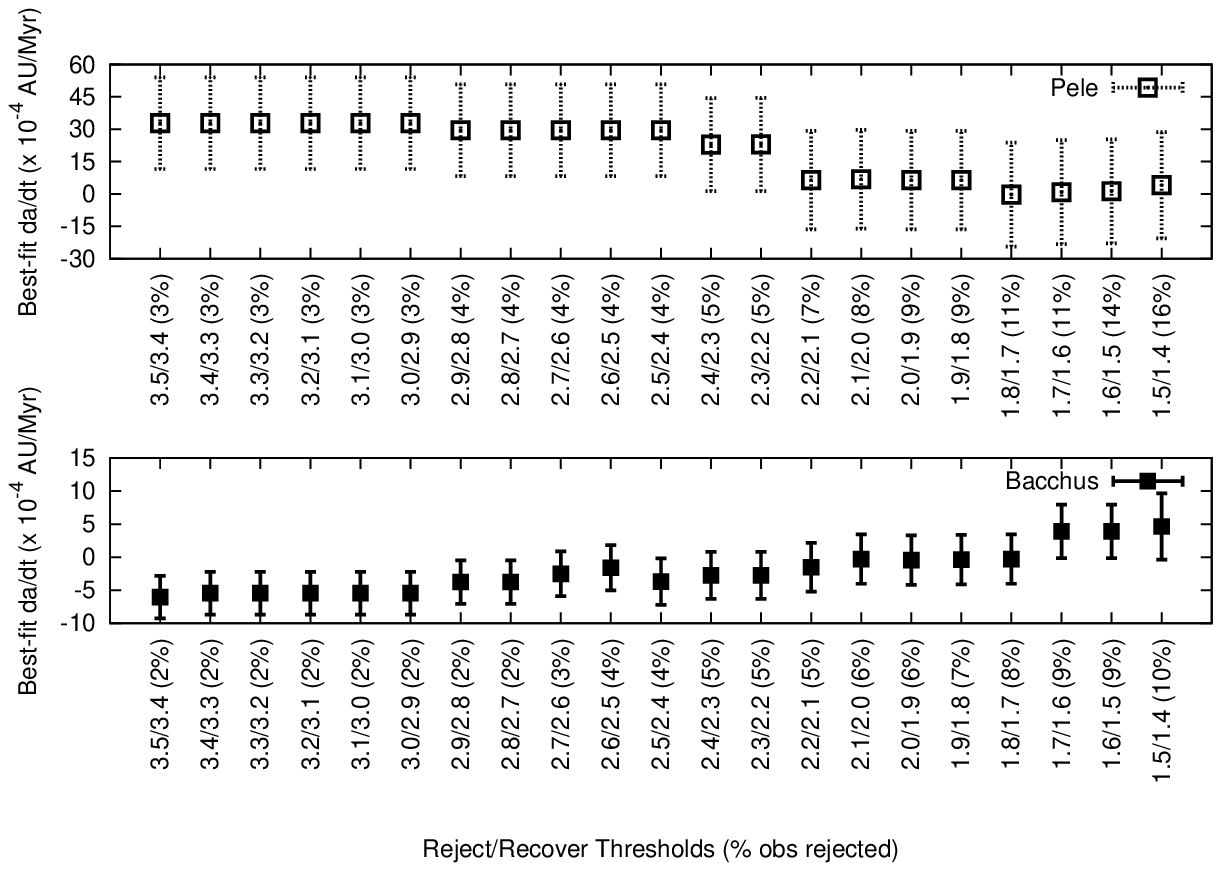}
	\end{center}
	\caption{Impact of different choices of reject/recover
          thresholds for the initial rejection step ($da/dt=0$) on the
          best-fit $da/dt$ values. Results from optical-only fits are
          shown with their $1\sigma$ error bars for two representative
          cases, (2202) Pele and (2063) Bacchus. Best-fit $da/dt$
          values are consistent with one another in the left half of
          the diagram.  Values to the right of 2.3/2.2 (Pele) and
          2.5/2.4 (Bacchus) have SNR less than unity and would not
          meet our selection criteria.  Our adopted reject/recover
          thresholds are $\sqrt{8}=2.828$ and $\sqrt{7}=2.646$.
	\label{fig:rejectrecover}
}
\end{figure}

We examined the impact of various choices of reject/recover thresholds
when rejecting outlier observations (Fig.~\ref{fig:rejectrecover}).
At moderate values of the rejection threshold (i.e.\ eliminating less
than $\sim$5\% of observations), best-fit values are consistent with
one another.  In this regime, results are fairly robust against the
choice of rejection thresholds.  However results do become sensitive
to rejection thresholds when a larger fraction of observations is
rejected.  As the reject/recover thresholds become more stringent,
astrometry with evidence of semi-major axis drift is preferentially
rejected, and the best-fit $da/dt$ values approach zero.  Our adopted
reject/recover thresholds ($\sqrt{8}$/$\sqrt{7}$) are stringent enough
that they eliminate obvious outliers, but not so stringent as to
suppress the Yarkovsky signal.  In 52 out of 54 cases, repeating the
outlier rejection step with the best-fit $da/dt$ value resulted in no
appreciable change to the result.

As a validation step, we compared the semi-major axis drift rates
obtained with our procedure (both optical-only and radar+optical) to
previously published values (Table~\ref{tab-chescompare}).  We found good
agreement for Golevka~\citep{ChesleyGolevka,ChesleyLCMfits} and
1992~BF~\citep{Vok1992BF}, and for most, but not all, NEAs included in
a similar study done by \citet{ChesleyLCMfits}.  The differences
between our results and those of \citet{ChesleyLCMfits} can probably
be attributed to our use of debiased data, of improved data weights,
and of longer observational arcs extending to 2012.  Eight objects
included in Table \ref{tab-chescompare} meet our selection criteria
for detailed analysis in the rest of this paper:
(1620) Geographos, (1685) Toro, (1862) Apollo, (1865) Cerberus,
 (2063) Bacchus, (2100) Ra-Shalom, (2340) Hathor,
and (152563) 1992 BF.

\begin{table}[h]
\rotate
\begin{center}
\caption{Comparison of our optical-only results to radar+optical (r+o)
  results and to the results of a previous study by
  \citet{ChesleyLCMfits}.\label{tab-chescompare} Best-fit $da/dt$
  values in units of 10$^{-4}$ AU/Myr and their one-sigma
  uncertainties are listed for optical-only and radar+optical
  observations.  Also shown is the root-mean-square (RMS) of weighted
  residuals for the gravity-only ($da/dt=0$) solution and for the
  non-zero $da/dt$ solution (RMS').  We restrict the radar analysis to
  those objects that have range measurements on at least two
  apparitions; this excludes (1685) Toro, (1865) Cerberus, (2063)
  Bacchus, (2340) Hathor, (85953) 1999 FK21, and (152563) 1992 BF.}
\begin{tabular}{lrrrrrrrrrr}
\tableline\tableline
NEA & \multicolumn{1}{r}{radar} & \multicolumn{1}{r}{radar} & \multicolumn{1}{r}{r+o} & \multicolumn{1}{r}{optical-only} & \multicolumn{1}{r}{r+o} & \multicolumn{1}{r}{\small Chesley 08}\\ 
                                     & RMS     & RMS'    & RMS'   & $da/dt$           & $da/dt$          & $da/dt$ \\
\tableline
(1620) Geographos                    & 0.393   & 0.356   & 0.55   &  $-2.43 \pm 0.7$  & $-2.52 \pm 0.6$   &  $-1.18 \pm 0.39$ \\
(1685) Toro                          & \nodata & \nodata & 0.51   &  $-1.40 \pm 0.7$  & \nodata           &  $-0.52 \pm 0.27$ \\
(1862) Apollo                        & 1.111   & 0.403   & 0.61   &  $-1.79 \pm 0.6$  & $-2.38 \pm 0.3$   &  $-2.44 \pm 0.26$ \\
(1865) Cerberus                      & \nodata & \nodata & 0.54   &  $-5.11 \pm 2.7$  & \nodata           &  $-7.80 \pm 2.28$ \\
(2063) Bacchus                       & \nodata & \nodata & 0.59   &  $-4.17 \pm 3.7$  & \nodata           & $-10.59 \pm 2.21$ \\
(2100) Ra-Shalom                     & 0.488   & 0.594   & 0.51   &  $-4.79 \pm 2.2$  & $-5.45 \pm 1.5$   &  $-7.09 \pm 0.88$ \\
(2340) Hathor                        & \nodata & \nodata & 0.67   & $-14.55 \pm 3.6$  & \nodata           & $-13.94 \pm 3.84$ \\
(6489) Golevka                       & 0.879   & 0.387   & 0.61   &  $-2.05 \pm 12.6$ & $-5.74 \pm 0.7$   &  $-6.39 \pm 0.44$ \\
(54509) YORP\tablenotemark{a}        & 0.796   & 0.260   & 0.55   & $-25.98 \pm 37.4$ & $-35.63\pm 10.5$  & $-25.12 \pm 6.18$ \\
(85953) 1999 FK21\tablenotemark{b}   & \nodata & \nodata & 0.56   & $-10.44 \pm 1.5$  & \nodata           & $-14.13 \pm 2.35$ \\
(101955) 1999 RQ36\tablenotemark{b}  & 15.694  & 0.127   & 0.39   & $-12.90 \pm 7.1$  & $-18.9 \pm 0.2$   & $-15.69 \pm 4.99$ \\
(152563) 1992 BF\tablenotemark{c}    & \nodata & \nodata & 0.60   & $-12.84 \pm 1.0$  & \nodata           & $-10.78 \pm 0.73$ \\
\tableline
\end{tabular}
\tablenotemark{a}{This object is in a Sun-Earth horseshoe
  orbit~\citep{tayl07}.}  \tablenotemark{b}{This object experiences
  perihelion precession of $\sim$16
  arcseconds/century~\citep{marg09iau261}.}\tablenotemark{c}{This
  object is the target of the OSIRIS-REx mission~\citep{ches12acm}.}
\tablenotemark{d}{Fits to this object use the astrometry corrections
  given in \citet{Vok1992BF} for the 1953 observations, which we did
  not subject to rejection.}
\end{center}
\end{table}

Several conclusions can be drawn from the data presented in Table
\ref{tab-chescompare}.  First, the RMS values indicate excellent fits
to the astrometry.  Second, the solutions with non-zero $da/dt$ values
provide a much better match to the radar data than the gravity-only
solutions, with typical RMS values decreasing by a factor of 2 or
more.  Third, radar+optical estimates have consistently lower error
bars than optical-only estimates, sometimes dramatically so, which is
typical in NEA studies.  Finally, there is a generally good agreement
between the optical-only $da/dt$ values and the radar+optical $da/dt$
values, indicating that the optical-only technique is a useful tool
that can be used even in the absence of radar data.

Drift rates for the 54 NEAs that pass our selection criteria are
presented in Table \ref{tbl-results} along with orbital elements and
physical properties. If an object has both a optical-only and a 
radar+optical value, we used the more accurate radar+optical value 
in the following figures and calculations (unless specified otherwise).
 We used Equation (\ref{equ:dadttyp}) with a
density of 1,200 kg m$^{-3}$ to compute efficiency factors $f_Y$ and
found that objects divided roughly into two groups.

\begin{figure}[ht]
	\begin{center}
		\includegraphics[scale=1.1]{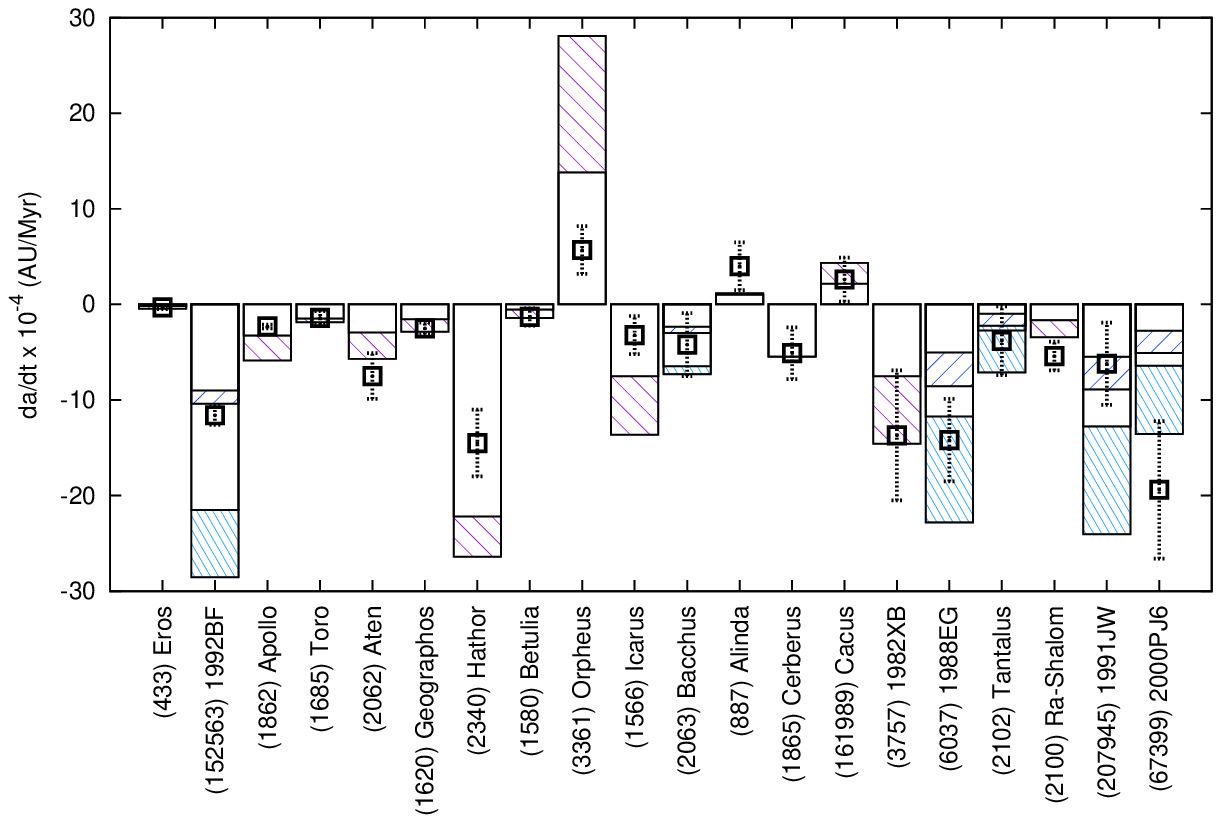}
	\end{center}
	\caption{ Measured and predicted drift values for 20 asteroids
          with Yarkovsky-dominated drifts, ordered by decreasing value
          of Yarkovsky sensitivity.  Best fits to optical-only data
          are shown as squares with dotted $1\sigma$ error bars.
          Shaded boxes show a range of predicted Yarkovsky rates
          representing different compositions (Table
          \ref{tab-thercon}). As predicted Yarkovsky values were
          calculated assuming $0^\circ$ or $180^\circ$ obliquity, the shaded
          boxes represent maximum drifts for the object. Therefore, a
          fit that lies between a shaded box and $da/dt=0$ is
          considered to have good agreement.  Objects with a single
          corresponding shaded box have a known diameter (Table
          \ref{tbl-results}). Objects with two shaded boxes did not
          have known diameters, and were modeled using diameters
          derived from assumed albedos (45\% in light blue, larger
          predicted drift magnitudes, and 5\% in dark blue, smaller
          predicted drift magnitudes). The vertical extents of the
          shaded boxes represent the range of compositional types
          described in Table \ref{tab-thercon}, with the larger
          absolute values representing the ``rubble pile" composition,
          and the lower absolute values representing the ``rock chunk"
          composition.  }
	\label{fig:defyark}
\end{figure}

\begin{figure}[ht]
	\begin{center}
		\includegraphics[scale=1.1]{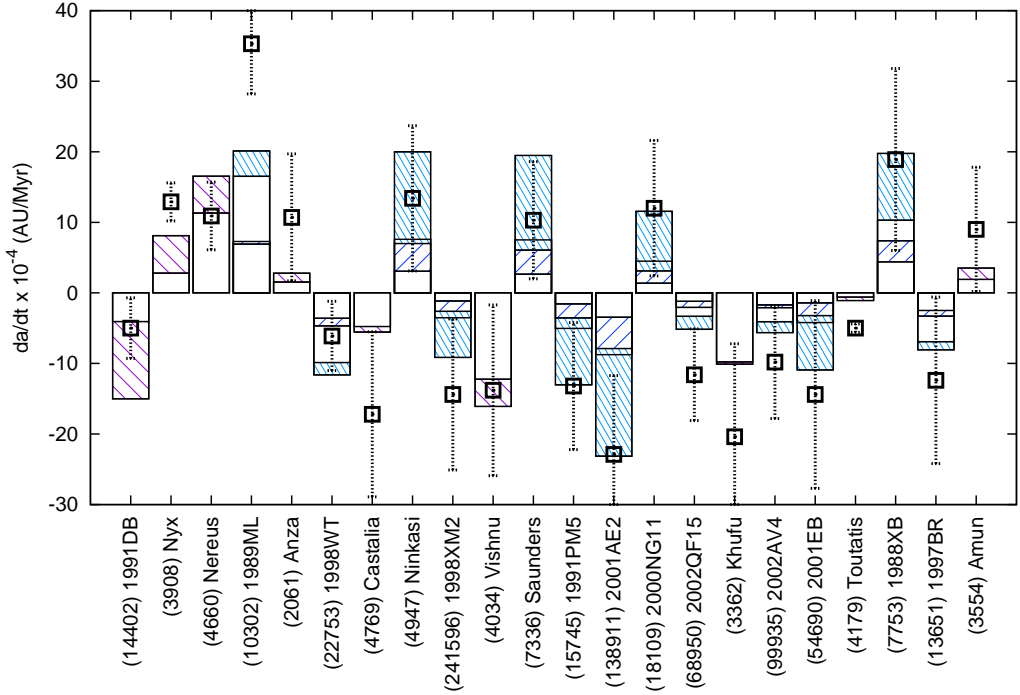}
	\end{center}
	\caption{Measured and predicted drift values for an additional
          22 asteroids with Yarkovsky-dominated drifts, ordered by
          decreasing value of Yarkovsky sensitivity.  Symbols are as
          in Fig.~\ref{fig:defyark}.  The observed rates for the
          majority of objects shown in this figure appear to exceed
          predicted values.  This is a consequence of the SNR $> 1$
          selection criterion which eliminates objects with lower
          $da/dt$ values.}
	\label{fig:defyarkb}
\end{figure}

\begin{figure}[ht]
	\begin{center}
		\includegraphics[scale=1.1]{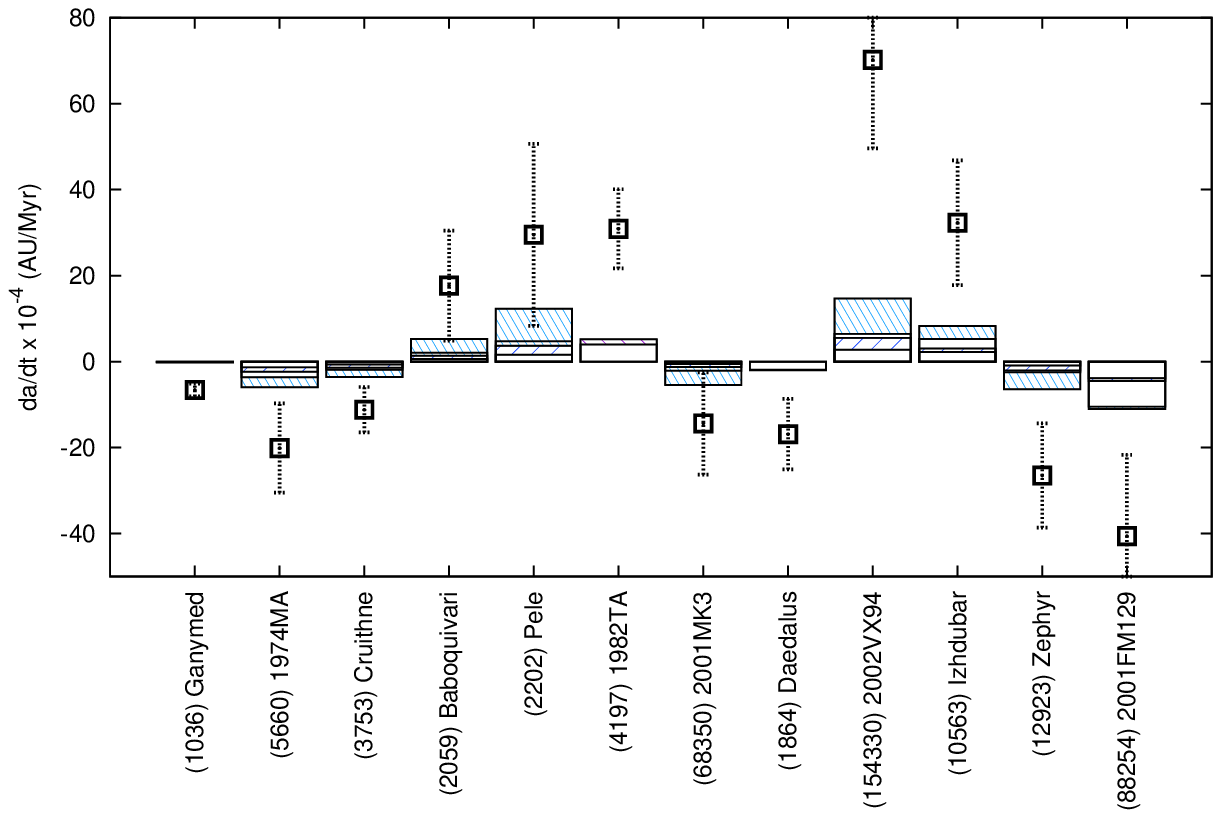}
	\end{center}
	\caption{Measured and predicted drift values for 12 asteroids
          with possible Yarkovsky-dominated drifts, defined as objects
          with Yarkovsky efficiency $f_Y$ exceeding 2 $\times
          10^{-5}$.  Symbols are as in Fig.~\ref{fig:defyark}.  Most
          objects in this figure have measured drifts that lie outside
          of the range of values expected on the basis of Yarkovsky
          models.  This could be due to inaccuracies in our knowledge
          of physical properties, faulty astrometry, or modeling
          errors.}
	\label{fig:possyark}
\end{figure}

In the first group of 42 objects with $f_Y \leq 2 \times 10^{-5}$,
most observed $da/dt$ values are consistent (within $1\sigma$) with
Yarkovsky predictions.  We refer to these objects as {\em
  Yarkovsky-dominated} (Figs.~\ref{fig:defyark} and
\ref{fig:defyarkb}).  In the second group of 12 objects with $f_Y > 2
\times 10^{-5} $ the observed $da/dt$ values are somewhat larger than
Yarkovsky predictions, but improvements in the knowledge of physical
properties or in Yarkovsky modeling could plausibly bring some of the
observed rates in agreement with predictions.  We refer to these
objects as {\em possibly Yarkovsky-dominated}
(Fig.~\ref{fig:possyark}).  

Figures~\ref{fig:defyark} and~\ref{fig:defyarkb} indicate that there
is generally agreement between observations and numerical estimates of
Yarkovsky drift rates for NEAs with $f_Y \le 2 \times 10^{-5}$.  These
data suggest that $f_Y \sim 10^{-5}$ represents a typical efficiency
for the Yarkovsky process. Predicted values are based on calculations
with obliquities of $0^\circ$ and $180^\circ$, therefore, observed rates that
are lower than predictions could still be due to the Yarkovsky effect.

The majority of objects in Fig.~\ref{fig:defyarkb} appear to exceed
predictions. This is a consequence of the SNR $> 1$ selection
criterion, as it eliminates objects with lower $da/dt$ values.  

On the basis of the entire sample of measured drifts for objects with
$s_Y > 2$, we can compute average properties for observed Yarkovsky
rates and efficiencies.  The mean, mean weighted by uncertainties,
median, and dispersion are shown in Table~\ref{tab-meanmedian}.  The
aggregate properties are comparable if we restrict objects to the
subset with SNR $ > $ 1, except for slightly increased $da/dt$ rates
(median rate of $\sim 12 \times 10^{-4}$ AU/Myr instead of $\sim 7
\times 10^{-4}$ AU/Myr), as expected.  The Yarkovsky process appears
to have an efficiency $f_Y$ of order $10^{-5}$, with a fairly small
dispersion.  Because the Yarkovsky efficiency scales with density
($f_Y|_{\rho}=f_Y|_{1,200} \times {\rho} / {1,200\ \rm{kg\ m}^{-3}}$)
some of the observed scatter is due to density variations.

\section{Discussion}

In this section we examine several consequences of our results.  First
we discuss how the Yarkovsky drifts can inform us about asteroid
physical properties, spin states, and trajectories. Then we discuss
binary asteroid (1862) Apollo and the curious case of asteroid (1036)
Ganymed.  Finally we discuss the possible mechanisms for non-Yarkovsky
driven rates, including association with meteoroid streams and rock
comet phenomenon.

\subsection{Yarkovsky-derived constraints on asteroid physical properties}
Because a clear connection exists between asteroid physical properties
and Yarkovsky drifts, we explored the constraints that can be placed
on bulk density and surface thermal conductivity for seven objects
with well-known diameters and (excepting one case) spin periods:
(1620) Geographos, (1862) Apollo, (2100) Ra-Shalom, (2062) Aten,
(2340) Hathor, (1566) Icarus, and (3361) Orpheus.
We compared the measured Yarkovsky rates to numerical estimates
obtained with a range of physical parameters. For these estimates, we
assumed a constant heat capacity $C = 500$ J kg$^{-1}$ K$^{-1}$ (Table
\ref{tab-thercon}) and a single value of the bulk density of the
surface $\rho_s=1,700$ kg~m$^{-3}$, but we explore a wide range of
bulk density and surface thermal conductivity values.  Because the
obliquities are uncertain or ambiguous in many cases, we chose to
illustrate outcomes for two obliquity values, typically 180$^\circ$
and 135$^\circ$.

\begin{figure}[ht]
	\begin{center}
		\includegraphics[scale=0.65]{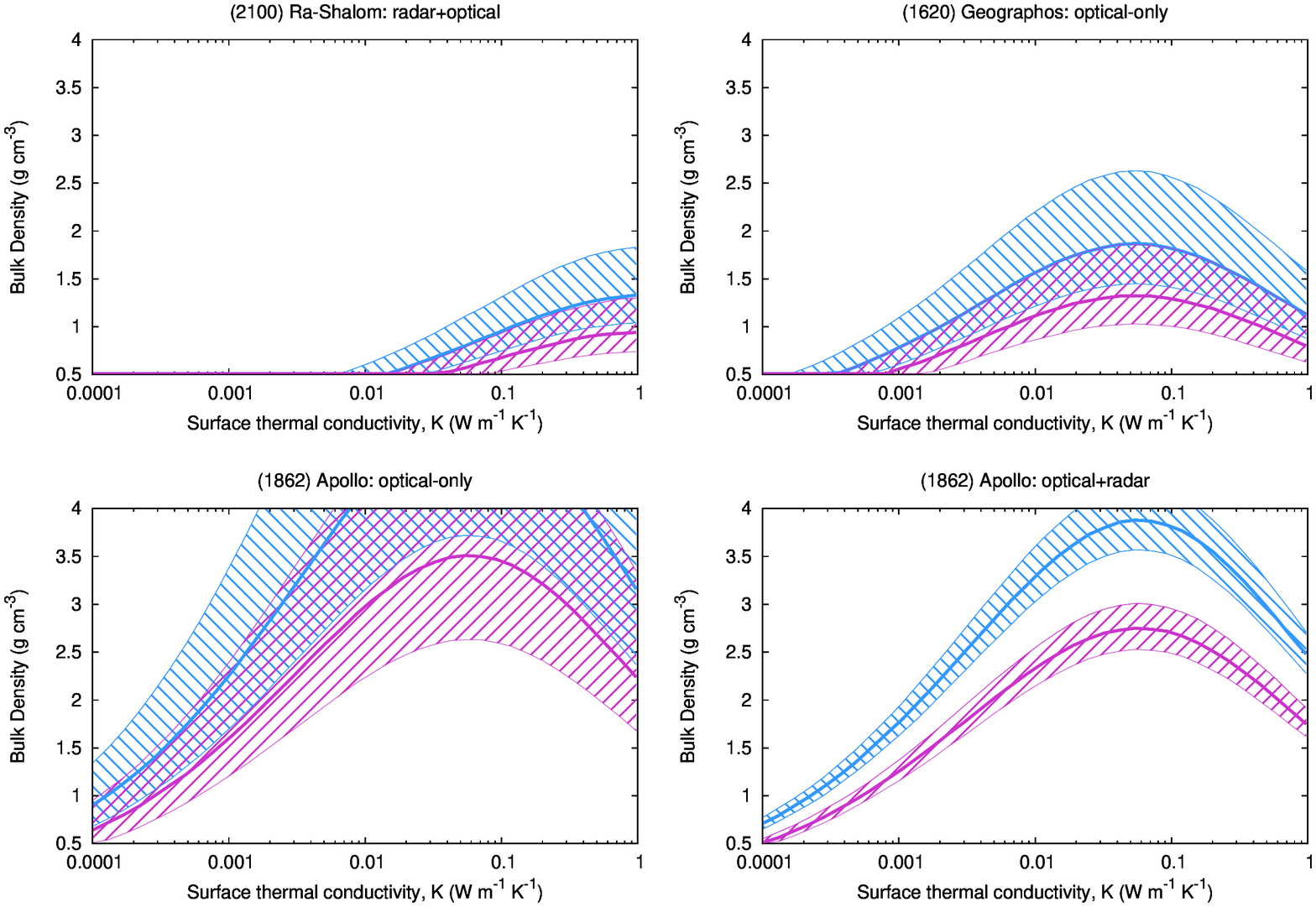}
	\end{center}
	\caption{Range of bulk densities and thermal conductivities of
          three Yarkovsky-dominated asteroids consistent with their
          observed $da/dt$ values.  Blue (top) solid line corresponds
          to values consistent with best-fit $da/dt$ and $180^\circ$
          obliquity, pink (lower) solid line corresponds to values
          consistent with best-fit $da/dt$ and $135^\circ$ obliquity.
          Dashed regions surrounding each solid line encompass the
          $1\sigma$ confidence limits on the corresponding $da/dt$
          determinations.  Not all values displayed in this K-$\rho$
          space are necessarily appropriate for asteroids.  Infrared
          observations suggest that (2100) Ra-Shalom has a thermal
          conductivity between 0.1 and 1 W~m$^{-1}$~K$^{-1}$
          \citep{DelboRaShalom,ShepardRaShalom}, consistent with the
          range suggested by our Yarkovsky rate determination.  For
          Apollo, we show results for both optical-only and
          radar+optical determinations.  The inclusion of radar data
          greatly reduces the error bars on the measured drift, and
          therefore the area of the shaded curves.  }
	\label{fig:rhoka}
\end{figure}

\begin{figure}[ht]
	\begin{center}
		\includegraphics[scale=0.65]{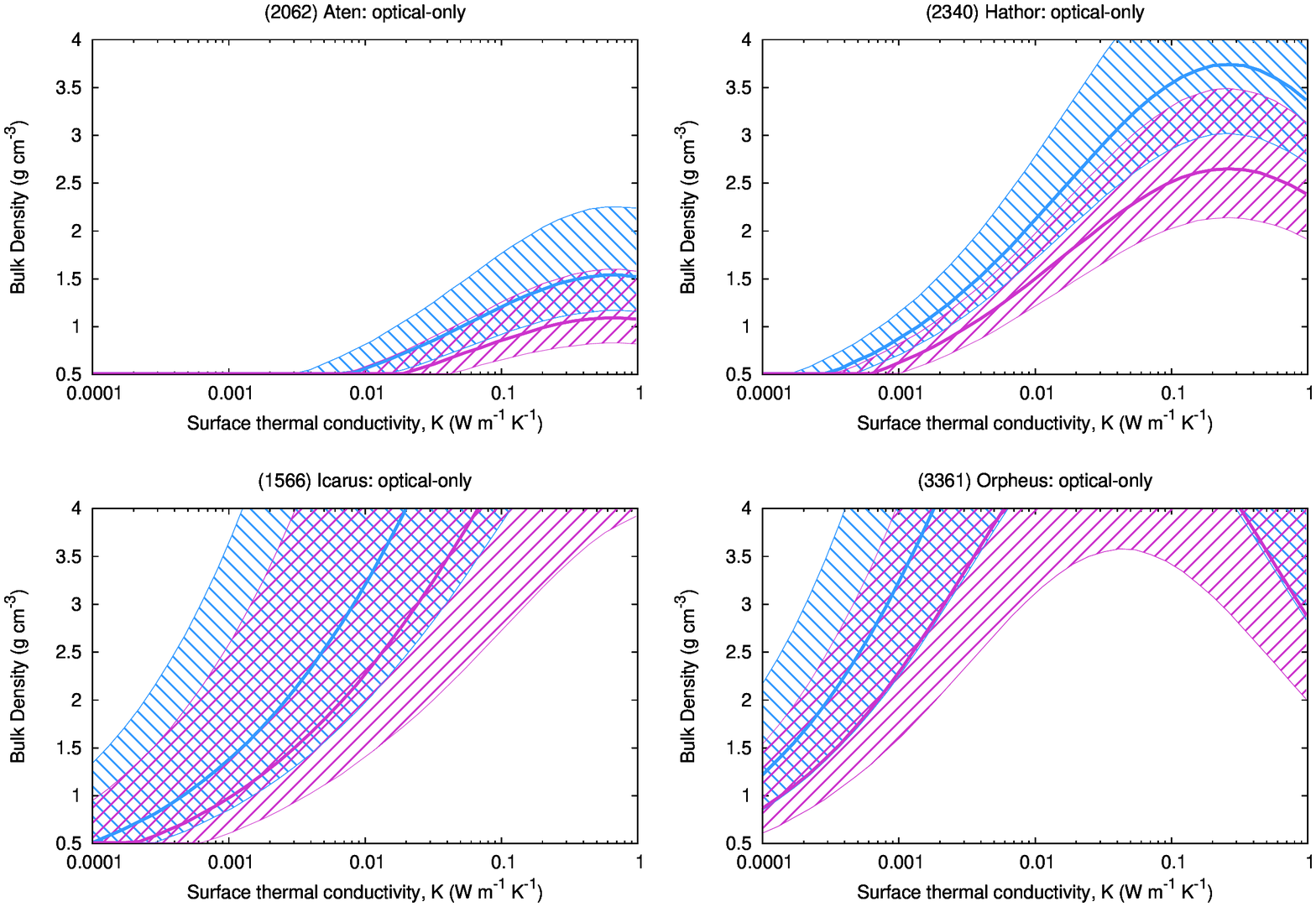}
	\end{center}
	\caption{Companion to Fig.~\ref{fig:rhoka}.  Range of bulk
          densities and thermal conductivities of three
          Yarkovsky-dominated asteroids consistent with their observed
          $da/dt$ values.  For (2062) Aten and (2340) Hathor, blue
          (top) solid line corresponds to values consistent with
          best-fit $da/dt$ and $180^\circ$ obliquity, pink (lower)
          solid line corresponds to values consistent with best-fit
          $da/dt$ and $135^\circ$ obliquity.  The constraints for
          (1566) Icarus suggest that it may have a lower obliquity
          than those assumed. (3361) Orpheus has a positive drift, so
          the blue (top) solid line corresponds to values consistent
          with best-fit $da/dt$ and $0^\circ$ obliquity, pink (lower)
          solid line corresponds to values consistent with best-fit
          $da/dt$ and $45^\circ$ obliquity.  Dashed regions
          surrounding each solid line encompass the $1\sigma$
          confidence limits on the corresponding $da/dt$
          determinations. Not all values displayed in this K-$\rho$
          space are necessarily appropriate for asteroids.  A period
          of 4.5 hours was assumed for (2340) Hathor, and a 0.15
          geometric albedo was assumed for (3361) Orpheus.  }
	\label{fig:rhokb}
\end{figure}

Our results are shown in Figures~\ref{fig:rhoka} and \ref{fig:rhokb},
which are similar to Fig. 4 in \citet{ChesleyGolevka}.  The shaded
range consistent with the $1\sigma$ confidence limits on $da/dt$
delineates the space of acceptable bulk densities and thermal
conductivities, assuming that the Yarkovsky effect is being modeled
correctly.  By acceptable, we mean consistent with observed $da/dt$
values, even though some of the $K-\rho$ values may not be appropriate
for asteroids.
 
Infrared observations indicate that (2100) Ra-Shalom has a thermal
conductivity between 0.1 and 1 W~m$^{-1}$~K$^{-1}$
\citep{DelboRaShalom,ShepardRaShalom}.  If we assume a minimum bulk
density of 1,500 kg~m$^{-3}$, this conductivity value is consistent with
the range suggested by our Yarkovsky rate determination.

If we make the same minimum density assumption for (1620) Geographos,
our measurements suggest that its surface thermal conductivity is
greater than 0.002 W~m$^{-1}$~K$^{-1}$.

For (1862) Apollo, we show the range of physical properties that are
consistent with both the optical-only fits and the radar+optical
fits. The precision of the radar measurements dramatically shrinks the
size of the measured error bars, with correspondingly tighter
constraints on density and surface thermal conductivity.  This example
illustrates that reliable obliquity determinations will be important
to extract physical properties from Yarkovsky rate determinations.

Our measurement of (2062) Aten's drift provides some useful
insights. If we assume that its bulk density exceeds $1,500$
kg~m$^{-3}$, then its surface thermal conductivity $K$ must exceed 0.3
W~m$^{-1}$~K$^{-1}$. Furthermore, if we assume that its bulk density
exceeds $1,600$ kg~m$^{-3}$, the $1\sigma$ confidence region on the
measured Yarkovsky drift suggests that its obliquity is between
$180^\circ-135^\circ$.

The Yarkovsky simulations for (2340) Hathor were computed with an
assumed spin period of 4.5 hours. If the actual period is longer, the
curves shown would shift to the left, and if the period is shorter,
the curves would shift to the right. Consequently, we cannot make
inferences about the $K$ value for this object until its spin period
is measured. However, looking at the height of the curve, and with an
assumption that the object's bulk density is greater than $1,500$
kg~m$^{-3}$, we can conclude that (2340) Hathor likely has an
obliquity lower than $180^\circ$.

The assumption of $135^\circ$ or $180^\circ$ obliquity for (1566)
Icarus restricts this object to low surface conductivity values and
low bulk density values, or high surface conductivity values and high
bulk density values. Although these obliquities do produce physically
plausible parameter combinations, it seems likely that the obliquity
for this object is $ \le 135^\circ$.

The curves for (3361) Orpheus were calculated with an assumed geometric albedo
of 0.15. As (3361) Orpheus has a positive $da/dt$ value, obliquities
were assumed to be $0^\circ$ and $45^\circ$. The curve representing an
obliquity equal to $0^\circ$ for this object requires very low ($<
0.002$ W~m$^{-1}$~K$^{-1}$) or very high ($> 0.7 $
W~m$^{-1}$~K$^{-1}$) surface thermal conductivity values for most
densities. A more
likely scenario is that this object has an obliquity $>0^\circ$, or
perhaps even $>45^\circ$.  An independent measurement of the obliquity
could be used to validate obliquity constraints derived from Yarkovsky
measurements.

\subsection{Yarkovsky rates and distribution of spin states}
\citet{LaSpinaNEA} and \citet{ChesleyLCMfits} examined the
predominance of retrograde spins and negative Yarkovsky drift rates
and concluded that they were consistent with the presumed delivery
method of NEAs from the main belt of asteroids.  The $\nu_6$ and 3:1
resonance regions deliver NEAs to near-Earth space \citep{bott02i}. A
main belt asteroid can arrive at the 3:1 resonance at 2.5 AU via a
positive (if it originates in the inner main belt) or negative (if it
originates in the outer main belt) Yarkovsky drift. However, a main
belt asteroid can only arrive at the $\nu_6$ resonance (at the inner
edge of the main belt) by way of a negative drift. According to
\citet{bott02i} and \citet{morb03}, $30\%-37\%$ of NEAs are
transported via the $\nu_6$ resonance, with the rest from other
resonances. The net result is a preference for retrograde spins.

An observational consequence of this process would be an excess of
retrograde rotators in the near-Earth asteroid
population. \citet{LaSpinaNEA} conducted a survey of 21 NEAs and found
the ratio of retrograde/prograde rotators to be $2.0^{+1}_{-0.7}$.

We note that out of the 42 Yarkovsky-dominated NEAs, 12 have a
positive $da/dt$ value. For this sample, our ratio of
retrograde/prograde rotators is $2.5 \pm 0.1$, similar to the value
found by \citet{LaSpinaNEA}.

\subsection{Impact of drift rates on asteroid trajectory predictions}
The semi-major axis drifts described in this paper affect NEA
trajectory predictions.  An order of magnitude estimate for the along
track displacement due to a non-zero $da/dt$ is given in
\citet{VokSmallNEAS}:
\begin{equation}
\Delta \rho \simeq 7 \dot{a}_4 (\Delta_{10} t)^2 a_{AU}^{-3/2}
\end{equation}
where $\Delta \rho$ is in units of km, $\dot{a}_4$ is $da/dt$ in
$10^{-4}$ AU/Myr, $\Delta_{10} t$ is the time difference between
observations in tens of years, and $a_{AU}$ is the semimajor axis of
the object in AU.  For instance, the estimated along-track
displacement due to the observed $da/dt$ for (1862) Apollo is 9~km
after 10 years.  Similarly, the estimated along-track displacement for
faster-moving (1864) Daedalus is 67 km after 10 years.

Our data indicate that (101955) 1999 RQ36, the target of the
OSIRIS-REx mission, has a measurable Yarkovsky drift of $(-18.9 \pm
0.2) \times 10^{-4}$ AU/Myr. Although it has a relatively short arc
(12 years) it has been observed three times by radar, allowing for an
accurate $da/dt$ measurement.  We estimated the along-track
displacement of (101955) 1999~RQ36 over the 6-month duration of the
OSIRIS-REx mission to be $0.3$ km, which will be easily detectable by
a radio science instrument.

\subsection{Binary asteroid (1862)  Apollo}
(1862) Apollo is a binary asteroid~\citep{iauc8627}.  Binary asteroids
present a unique opportunity for the determination of physical
parameters. If mass and density can be measured from the binary orbit
and component sizes, the Yarkovsky constraint on thermal conductivity
can become much more meaningful.  If the orientation of the plane of
the mutual orbital can be measured, a plausible obliquity can be
assumed, which makes the constraints on thermal properties tighter
still.  In some cases, actual obliquity measurements can be obtained
from shape modeling efforts.

\citet{Yeomanscomet,YeomansErratum} identified a non-gravitational
perturbation acting on the orbit of (1862) Apollo, but was not able to
determine a drift magnitude.  To $1\sigma$ our observed $da/dt$ value
for (1862) Apollo agrees with our Yarkovsky predictions.

\subsection{The curious case of (1036) Ganymed}
(1036) Ganymed has by far the largest Yarkovsky efficiency value ($f_Y
\sim 15 \times 10^{-5}$) among the objects presented in Table
\ref{tbl-results}.  
With a nominal value of $\sim -7 \times 10^{-4}$ AU/Myr, the measured
$da/dt$ value is comparable to that of other NEAs.  Combined with
Ganymed's large diameter estimate ($\sim 32$ km based on IRAF
measurements), this Yarkovsky rate results in an unusually high $f_Y$
value.

How can this anomaly be explained?  One possibility is that some of
the early astrometry, dating back to 1924, is erroneous.  This could
be due to measurement errors, timing errors, bias errors, or reference
frame conversion errors.  We evaluated the semi-major axis drift with
various subsets of the available astrometry and found values ranging
between $-3 \times 10^{-4}$ and $-8 \times 10^{-4}$ AU/Myr.  On that
basis we modified the adopted uncertainties for this object, and our
preferred value is $(-6.62_{-1.4}^{+3.6}) \times 10^{-4}$ AU/Myr.
Doing so does not eliminate the possibility of systematic bias in the
astrometry, and we are still left with anomalously high $f_Y$ values.

Another possibility is that the diameter of Ganymed, an S-type
asteroid, is much smaller than reported.  This seems unlikely
considering the more recent WISE albedo measurement of $p_V = 0.212$
\citep{Masiero12} which suggests a diameter of $\sim 36$ km.

If Ganymed's bulk density was especially low, a higher than usual
$f_Y$ value would be expected, but this would likely explain a factor
of 2 or 3 at most, and would not explain the anomalous value.

Perhaps Ganymed departs significantly from a spherical shape, with an
effective diameter and mass that are much smaller than those implied
by the diameter values reported in the literature.  The relatively low
lightcurve amplitudes do not seem to support such an argument, unless
the asteroid is particularly oblate.  In that case one could plausibly
arrive at volume and mass estimates that are off by a factor of 5-10.

If we can rule out these possibilities (i.e.\ Ganymed is roughly
spherical with no substantial concavities, its diameter estimate is
reasonably accurate, and the early astrometry can be trusted), and if
no other modeling error can be identified, then we would be compelled
to accept an anomalously high Yarkovsky efficiency for this object.

\subsection{Non-Yarkovsky processes}

In the course of our study we observed drift values that cannot be
accounted for easily by Yarkovsky drift, because they considerably
exceed the predicted Yarkovsky rates.  In most cases, these can be
attributed to poor sensitivity to Yarkovsky influences
(Fig.~\ref{fig-xmas}).  Therefore, the high rates can generally be
safely discarded.  In other cases, the high rates may be due to
erroneous optical astrometry or mismodeling of asteroid-asteroid
perturbations.  However we cannot entirely rule out the possibility
that some of the high drift rates are secure and will be confirmed by
further observation and analysis.  If the high rates cannot be
ascribed to poor Yarkovsky sensitivity or faulty astrometry, one would
need to invoke other non-gravitational forces.

One possibility is that orbits are perturbed when NEAs are losing gas
or dust in an anisotropic manner.
To estimate a rough rate of mass loss that would be needed to account
for the drifts measured, we used the basic thrust equation
\begin{equation}
F=qV_{e}
\end{equation}
where $F$ is the force, $q$ is the rate at which the mass departs the
asteroid, and $V_{e}$ is the ejection speed.  For an asteroid of mass
$m$ this yields
\begin{equation}
a_{\rm mass\ loss}=\frac{qV_{e}}{m}
\end{equation}
which can be incorporated into Gauss' form of Lagrange's planetary
equations \citep{Danby} as an acceleration aligned with the velocity
of the object. The dependence of the force on heliocentric distance
$r$ is not known precisely; we assumed $F \propto r^{-2}$, similar to
the Yarkovsky dependence, for simplicity, and because the amount of
outgassing likely scales with the amount of incident radiation (as in
Fig.~4 of \citet{Delsemme}).  We assumed $V_e=1.5$ m s$^{-1}$, the
value derived by \citet{JewittElst} for 133P/Elst-Pizarro, and we
assumed that the mass is departing in the optimal thrust direction.

We quantified the mass loss rates needed to produce the observed
drifts of NEAs with the highest Yarkovsky efficiencies.  We estimated
a rate of 0.16~kg s$^{-1}$ for (154330) 2002 VX94 and 2.3 kg s$^{-1}$
for (7889) 1994 LX.
Although these estimates represent the minimum amount of mass loss
necessary to account for the observed drifts (if due to mass loss),
they are smaller than typical levels from comets.  Comets have mass
loss rates that span a wide range of values.  On the high side a rate
of $ 2 \times 10^6$ kg s$^{-1}$ was estimated for Hale-Bopp
\citep{JewittHaleBopp}.  On the low side \citet{Ishiguro2007169}
measured mass loss rates for three comets, averaged over their orbits:
2P/Encke ($48 \pm 20$ kg s$^{-1}$), 22P/Kopff ($17 \pm 3$ kg
s$^{-1}$), and 65P/Gunn ($27 \pm 9$ kg s$^{-1}$).  Mass loss rates of
active asteroids have been estimated to be in the range from $\le
0.04$ kg s$^{-1}$ (113P/Elst-Pizarro) to $\le 150$ kg s$^{-1}$
(107P/Wilson-Harrington) \citep{Jewitt12}.

Mass loss does not seem to be a viable mechanism to explain the
semi-major axis drift rate of (1036) Ganymed, as it would require a
minimum mass loss rate of $\sim$2,500 kg s$^{-1}$.  This would presumably
have left detectable observational signatures, which have not been
reported to date.

We explore a couple of possibilities for mass loss mechanisms that
could cause semi-major axis drifts.

\subsubsection{Associations with meteoroid streams}

To our knowledge, (433) Eros, (1566) Icarus, (1620) Geographos, (1685)
Toro, (1862) Apollo, and 1982 TA are the only objects in our sample
to have been associated with a meteoroid
stream. \citet{SekaninaMeteor} found a weak correlation between the
first five objects and various streams using the ``dissimilarity
criterion". However, this metric was later described as not
convincing by \citet{Jenniskens}, and current literature does not
support such associations.  In our results, Apollo shows good
agreement with Yarkovsky predictions, with $f_Y = 0.25\times 10^{-5}$.
The Yarkovsky force is therefore a plausible cause of Apollo's
observed semi-major axis drift.

\subsubsection{Rock comet phenomenon}
The brightening of (3200) Phaethon, the parent body of the Geminid
meteor shower, has been attributed to a ``rock comet'' phenomenon
\citep{JewittPhaethon}.  With a perihelion at 0.14 AU, (3200)
Phaethon's surface temperatures have been estimated by
\citet{JewittPhaethon} to be in the range $746 < T < 1050$ K.  The
authors propose that these high surface temperatures could create
thermal gradients in the body, resulting in thermal fracturing that
would release dust.  The resulting mass loss would
affect the orbit.  The combination of mass loss due to decomposing
hydrated minerals and thermal fracturing led the authors to term
(3200) Phaethon a ``rock comet''.  A moderate amount ($\sim$ 1 kg s$^{-1}$)
of mass lost in an anisotropic manner by ``rock comets'' could explain
the observed semi-major axis drift rates.

\section{Conclusions}

Modeling of the Yarkovsky effect is needed to improve trajectory
predictions of near-Earth asteroids and to refine our understanding of
the dynamics of small bodies.  Using fits to astrometric data, we
identified semi-major axis drifts in 54 NEAs, 42 of which show good
agreement with numerical estimates of Yarkovsky drifts, indicating
that they are likely Yarkovsky-dominated.  These objects exhibit
Yarkovsky efficiencies of $\sim$10$^{-5}$, where the efficiency
describes the ratio of the change in orbital energy to incident solar
radiation energy. 12 objects in our sample have drifts that exceed
nominal Yarkovsky predictions and are labeled possibly
Yarkovsky-dominated.  Improvements in the knowledge of physical
properties or in thermal modeling could bring these drift rates in
better agreement with results from numerical models.  However, if the
high rates are confirmed by additional observations and analysis, they
would be indicative of the presence of other non-gravitational forces,
such as that resulting from a loss of mass.


\acknowledgments

None of this work would have been possible 
 without the availability of the
OrbFit software package (available at
\url{http://adams.dm.unipi.it/orbfit/ }).  

We are grateful to The Minor Planet Center and all astronomers who
submitted data to the Minor Planet Center.

We thank Ned Wright (UCLA) for insightful comments.

CN and JLM were partially funded by NSF Planetary Astronomy grants
AST-0929830 and AST-1109772.  Part of this research was conducted at
the Jet Propulsion Laboratory, California Institute of Technology,
under a contract with the National Aeronautics and Space
Administration.  DV was partially supported by the Czech Grant Agency
(grant 205/08/0064) and Research Program MSM0021620860 of the Czech
Ministry of Education.

\bibliographystyle{apj}

\begin{thebibliography}{46}
\expandafter\ifx\csname natexlab\endcsname\relax\def\natexlab#1{#1}\fi

\bibitem[{{Bottke} {et~al.}(2002{\natexlab{a}}){Bottke}, {Morbidelli},
  {Jedicke}, {Petit}, {Levison}, {Michel}, \& {Metcalfe}}]{bott02i}
{Bottke}, W.~F., {Morbidelli}, A., {Jedicke}, R., {Petit}, J., {Levison},
  H.~F., {Michel}, P., \& {Metcalfe}, T.~S. 2002{\natexlab{a}}, Icarus, 156,
  399

\bibitem[{{Bottke} {et~al.}(2002{\natexlab{b}}){Bottke}, {Vokrouhlick{\'y}},
  {Rubincam}, \& {Bro\v{z}}}]{BottkeYarkAstIII}
{Bottke}, Jr., W.~F., {Vokrouhlick{\'y}}, D., {Rubincam}, D.~P., \& {Bro\v{z}},
  M. 2002{\natexlab{b}}, Asteroids III, 395

\bibitem[{{Bottke} {et~al.}(2006){Bottke}, {Vokrouhlick{\'y}}, {Rubincam}, \&
  {Nesvorn{\'y}}}]{2006AREPS}
{Bottke}, Jr., W.~F., {Vokrouhlick{\'y}}, D., {Rubincam}, D.~P., \&
  {Nesvorn{\'y}}, D. 2006, Annual Review of Earth and Planetary Sciences, 34,
  157

\bibitem[{{Bowell} {et~al.}(1989){Bowell}, {Hapke}, {Domingue}, {Lumme},
  {Peltoniemi}, \& {Harris}}]{bowe89}
{Bowell}, E., {Hapke}, B., {Domingue}, D., {Lumme}, K., {Peltoniemi}, J., \&
  {Harris}, A.~W. 1989, in Asteroids II, 524--556

\bibitem[{{Chamberlin}(2008)}]{SmallBodyDatabase}
{Chamberlin}, A.~B. 2008, Bulletin of the American Astronomical Society, 40

\bibitem[{Chesley {et~al.}({2003})Chesley, Ostro, Vokrouhlick\'{y}, \v{C}apek,
  Giorgini, Nolan, Margot, Hine, Benner, \& Chamberlin}]{ChesleyGolevka}
Chesley, S., Ostro, S., Vokrouhlick\'{y}, D., \v{C}apek, D., Giorgini, J.,
  Nolan, M., Margot, J.~L., Hine, A., Benner, L., \& Chamberlin, A. {2003},
  {Science}, {302}, 1739

\bibitem[{{Chesley}(2006)}]{ChesleyHazard}
{Chesley}, S.~R. 2006, Asteroids, Comets, and Meteors conference abstracts

\bibitem[{Chesley {et~al.}({2010})Chesley, Baer, \& Monet}]{ChesleyBias}
Chesley, S.~R., Baer, J., \& Monet, D.~G. {2010}, {Icarus}, {210}, 158

\bibitem[{{Chesley} {et~al.}(2008){Chesley}, {Vokrouhlick{\'y}}, {Ostro},
  {Benner}, {Margot}, {Matson}, {Nolan}, \& {Shepard}}]{ChesleyLCMfits}
{Chesley}, S.~R., {Vokrouhlick{\'y}}, D., {Ostro}, S.~J., {Benner}, L.~A.~M.,
  {Margot}, J.~L., {Matson}, R.~L., {Nolan}, M.~C., \& {Shepard}, M.~K. 2008,
  LPI Contributions, 1405, 8330

\bibitem[{Chesley {et~al.}(2012)}]{ches12acm}
Chesley, S.~R. {et~al.} 2012, Asteroids, Comets, Meteors conference abstracts

\bibitem[{{Danby}(1992)}]{Danby}
{Danby}, J.~M.~A. 1992, {Fundamentals of Celestial Mechanics} (Willmann-Bell,
  Inc)

\bibitem[{Delb{\'o} {et~al.}({2007})Delb{\'o}, dell'Oro, Harris, Mottola, \&
  Mueller}]{Delbo07ThermInertYark}
Delb{\'o}, M., dell'Oro, A., Harris, A.~W., Mottola, S., \& Mueller, M. {2007},
  {Icarus}, {190}, 236

\bibitem[{{Delb{\'o}} {et~al.}(2003){Delb{\'o}}, {Harris}, {Binzel}, {Pravec},
  \& {Davies}}]{DelboRaShalom}
{Delb{\'o}}, M., {Harris}, A.~W., {Binzel}, R.~P., {Pravec}, P., \& {Davies},
  J.~K. 2003, Icarus, 166, 116

\bibitem[{{Delsemme}(1982)}]{Delsemme}
{Delsemme}, A.~H. 1982, in Comets, ed. {L.~L.~Wilkening}, 85--130

\bibitem[{{Giorgini} {et~al.}(2008){Giorgini}, {Benner}, {Ostro}, {Nolan}, \&
  {Busch}}]{gior08}
{Giorgini}, J.~D., {Benner}, L.~A.~M., {Ostro}, S.~J., {Nolan}, M.~C., \&
  {Busch}, M.~W. 2008, \icarus, 193, 1

\bibitem[{{Giorgini} {et~al.}(2002){Giorgini}, {Ostro}, {Benner}, {Chodas},
  {Chesley}, {Hudson}, {Nolan}, {Klemola}, {Standish}, {Jurgens}, {Rose},
  {Chamberlin}, {Yeomans}, \& {Margot}}]{gior02}
{Giorgini}, J.~D., {Ostro}, S.~J., {Benner}, L.~A.~M., {Chodas}, P.~W.,
  {Chesley}, S.~R., {Hudson}, R.~S., {Nolan}, M.~C., {Klemola}, A.~R.,
  {Standish}, E.~M., {Jurgens}, R.~F., {Rose}, R., {Chamberlin}, A.~B.,
  {Yeomans}, D.~K., \& {Margot}, J.~L. 2002, Science, 296, 132

\bibitem[{{Goldreich} \& {Sari}(2009)}]{GoldreichSari}
{Goldreich}, P. \& {Sari}, R. 2009, The Astrophysical Journal, 691, 54

\bibitem[{{Hsieh} {et~al.}(2004){Hsieh}, {Jewitt}, \&
  {Fern{\'a}ndez}}]{JewittElst}
{Hsieh}, H.~H., {Jewitt}, D.~C., \& {Fern{\'a}ndez}, Y.~R. 2004, The
  Astronomical Journal, 127, 2997

\bibitem[{{H\"{u}tter} \& {K\"{o}mle}(2008)}]{HutterRad}
{H\"{u}tter}, E.~S. \& {K\"{o}mle}, N.~I. 2008, in 5th European
  Thermal-Sciences Conference, The Netherlands

\bibitem[{Ishiguro {et~al.}(2007)Ishiguro, Sarugaku, Ueno, Miura, Usui, Chun,
  \& Kwon}]{Ishiguro2007169}
Ishiguro, M., Sarugaku, Y., Ueno, M., Miura, N., Usui, F., Chun, M.-Y., \&
  Kwon, S.~M. 2007, Icarus, 189, 169

\bibitem[{{Jenniskens}(2008)}]{Jenniskens}
{Jenniskens}, P. 2008, Earth Moon and Planets, 102, 505

\bibitem[{{Jewitt}(2012)}]{Jewitt12}
{Jewitt}, D. 2012, The Astronomical Journal, 143, 66

\bibitem[{{Jewitt} \& {Li}(2010)}]{JewittPhaethon}
{Jewitt}, D. \& {Li}, J. 2010, The Astronomical Journal, 140, 1519

\bibitem[{{Jewitt} \& {Matthews}(1999)}]{JewittHaleBopp}
{Jewitt}, D. \& {Matthews}, H. 1999, The Astronomical Journal, 117, 1056

\bibitem[{{Konopliv} {et~al.}(2011){Konopliv}, {Asmar}, {Folkner}, {Karatekin},
  {Nunes}, {Smrekar}, {Yoder}, \& {Zuber}}]{kono11}
{Konopliv}, A.~S., {Asmar}, S.~W., {Folkner}, W.~M., {Karatekin}, {\"O}.,
  {Nunes}, D.~C., {Smrekar}, S.~E., {Yoder}, C.~F., \& {Zuber}, M.~T. 2011,
  \icarus, 211, 401

\bibitem[{La~Spina {et~al.}({2004})La~Spina, Paolicchi, Kryszczynska, \&
  Pravec}]{LaSpinaNEA}
La~Spina, A., Paolicchi, P., Kryszczynska, A., \& Pravec, P. {2004}, {Nature},
  {428}, 400

\bibitem[{{Margot} \& {Giorgini}(2010)}]{marg09iau261}
{Margot}, J.~L. \& {Giorgini}, J.~D. 2010, in IAU Symposium, ed.
  {S.~A.~Klioner, P.~K.~Seidelmann, \& M.~H.~Soffel}, Vol. 261, 183--188

\bibitem[{{Masiero} {et~al.}(2012){Masiero}, {Mainzer}, {Grav}, {Bauer},
  {Wright}, {McMillan}, {Tholen}, \& {Blain}}]{Masiero12}
{Masiero}, J.~R., {Mainzer}, A.~K., {Grav}, T., {Bauer}, J.~M., {Wright},
  E.~L., {McMillan}, R.~S., {Tholen}, D.~J., \& {Blain}, A.~W. 2012, The
  Astrophysical Journal, 749, 104

\bibitem[{Milani {et~al.}({2009})Milani, Chesley, Sansaturio, Bernardi,
  Valsecchi, \& Arratia}]{MilaniImpact1999RQ36}
Milani, A., Chesley, S.~R., Sansaturio, M.~E., Bernardi, F., Valsecchi, G.~B.,
  \& Arratia, O. {2009}, {Icarus}, {203}, 460

\bibitem[{{Milani} \& {Gronchi}(2009)}]{MilaniOrbitBook}
{Milani}, A. \& {Gronchi}, G. 2009, {Theory of Orbit Determination} (Cambridge
  University Press)

\bibitem[{{Morbidelli} \& {Vokrouhlick{\'y}}(2003)}]{morb03}
{Morbidelli}, A. \& {Vokrouhlick{\'y}}, D. 2003, \icarus, 163, 120

\bibitem[{Opeil {et~al.}(2010)Opeil, Consolmagno, \& Britt}]{OpeilTherCon}
Opeil, C., Consolmagno, G., \& Britt, D. 2010, Icarus, 208, 449

\bibitem[{{Ostro} {et~al.}(2005){Ostro}, {Benner}, {Giorgini}, {Nolan}, {Hine},
  {Howell}, {Margot}, {Magri}, \& {Shepard}}]{iauc8627}
{Ostro}, S.~J., {Benner}, L.~A.~M., {Giorgini}, J.~D., {Nolan}, M.~C., {Hine},
  A.~A., {Howell}, E.~S., {Margot}, J.~L., {Magri}, C., \& {Shepard}, M.~K.
  2005, IAU Circular, 8627

\bibitem[{Pravec \& Harris({2000})}]{PravecFastSlow}
Pravec, P. \& Harris, A. {2000}, {Icarus}, {148}, 12

\bibitem[{Pravec \& Harris(2007)}]{PravecHarrisIcarus}
Pravec, P. \& Harris, A.~W. 2007, Icarus, 190, 250

\bibitem[{Press {et~al.}(1992)Press, Teukolsky, Vetterling, \&
  Flannery}]{NumericalRec}
Press, W.~H., Teukolsky, S.~A., Vetterling, W.~T., \& Flannery, B.~P. 1992,
  Numerical Recipes in C (2nd ed.): The Art of Scientific Computing (New York,
  NY, USA: Cambridge University Press)

\bibitem[{{Sekanina}(1976)}]{SekaninaMeteor}
{Sekanina}, Z. 1976, Icarus, 27, 265

\bibitem[{{Shepard} {et~al.}(2008){Shepard}, {Clark}, {Nolan}, {Benner},
  {Ostro}, {Giorgini}, {Vilas}, {Jarvis}, {Lederer}, {Lim}, {McConnochie},
  {Bell}, {Margot}, {Rivkin}, {Magri}, {Scheeres}, \&
  {Pravec}}]{ShepardRaShalom}
{Shepard}, M.~K., {Clark}, B.~E., {Nolan}, M.~C., {Benner}, L.~A.~M., {Ostro},
  S.~J., {Giorgini}, J.~D., {Vilas}, F., {Jarvis}, K., {Lederer}, S., {Lim},
  L.~F., {McConnochie}, T., {Bell}, J., {Margot}, J.~L., {Rivkin}, A., {Magri},
  C., {Scheeres}, D., \& {Pravec}, P. 2008, Icarus, 193, 20

\bibitem[{{Sitarski}(1992)}]{sitarskiIcarus}
{Sitarski}, G. 1992, The Astronomical Journal, 104, 1226

\bibitem[{{Sitarski}(1998)}]{sitarskiToutatis}
---. 1998, Acta Astronomica, 48, 547

\bibitem[{{Taylor} {et~al.}(2007){Taylor}, {Margot}, {Vokrouhlick{\'y}},
  {Scheeres}, {Pravec}, {Lowry}, {Fitzsimmons}, {Nolan}, {Ostro}, {Benner},
  {Giorgini}, \& {Magri}}]{tayl07}
{Taylor}, P.~A., {Margot}, J.~L., {Vokrouhlick{\'y}}, D., {Scheeres}, D.~J.,
  {Pravec}, P., {Lowry}, S.~C., {Fitzsimmons}, A., {Nolan}, M.~C., {Ostro},
  S.~J., {Benner}, L.~A.~M., {Giorgini}, J.~D., \& {Magri}, C. 2007, Science,
  316, 274

\bibitem[{Vokrouhlick\'{y} {et~al.}({2008})Vokrouhlick\'{y}, Chesley, \&
  Matson}]{Vok1992BF}
Vokrouhlick\'{y}, D., Chesley, S.~R., \& Matson, R.~D. {2008}, {Astronomical
  Journal}, {135}, 2336

\bibitem[{{Vokrouhlick{\'y}} {et~al.}(2000){Vokrouhlick{\'y}}, {Milani}, \&
  {Chesley}}]{VokSmallNEAS}
{Vokrouhlick{\'y}}, D., {Milani}, A., \& {Chesley}, S.~R. 2000, Icarus, 148,
  118

\bibitem[{{Yeomans}(1991)}]{Yeomanscomet}
{Yeomans}, D.~K. 1991, The Astronomical Journal, 101, 1920

\bibitem[{Yeomans(1992)}]{YeomansErratum}
Yeomans, D.~K. 1992, The Astronomical Journal, 104, 1266

\bibitem[{{Ziolkowski}(1983)}]{Ziol}
{Ziolkowski}, K. 1983, in Asteroids, Comets, and Meteors, ed.
  {C.-I.~Lagerkvist, H.~Rickman}, 171--174

\end{thebibliography}


\clearpage
\begin{deluxetable}{rlrrrrrrrrrrrrrr}

\tabletypesize{\scriptsize}
\rotate
\tablecaption{Semi-major Axis Drift Rates\label{tbl-results}}
\tablewidth{0pt}
\tablehead{
\colhead{NEA}   &\colhead{ }   &   \colhead{$a$}   &   \colhead{$e$}   &   \colhead{$i$}  &   
\colhead{$D$} & \colhead{$P$} & \colhead{$p_V$} & \colhead{Arc}  &\colhead{$(da/dt)_{\rm o}$}  &   \colhead{$1\sigma_{\rm}$} &
\colhead{$(da/dt)_{\rm r+o}$}  &   \colhead{$1\sigma_{\rm}$} & \colhead{SNR}& \colhead{$s_Y$} & \colhead{$f_Y  $}   \\
 &    &   (AU)   &     &   (deg)   &   (km)   &   (h)   &       &     &   \multicolumn{2}{c}{(10$^{-4}$ AU/Myr)}   &   \multicolumn{2}{c}{(10$^{-4}$ AU/Myr)}  & & & $  \times10^{-5}$  
}
\startdata
    (433) &  Eros              & 1.46 &  0.22 & 10.83 & 16.84          &    5.270  &    0.25  & 1893-2012 &   -0.3 &   0.2 &  \nodata & \nodata &  1.81 & 70.56 &  0.38\\ 
 (152563) &  1992 BF           & 0.91 &  0.27 &  7.25 &  0.42$\dagger$ &   \nodata &  \nodata & 1992-2011 &  -11.6 &   1.0 &  \nodata & \nodata & 11.26 & 40.28 &  0.37\\ 
   (1862) &  Apollo            & 1.47 &  0.56 &  6.35 &  1.50          &    3.065  &    0.25  & 1957-2012 &   -1.8 &   0.6 &   -2.3   &     0.2 & 11.50 & 36.11 &  0.23\\ 
   (1685) &  Toro              & 1.37 &  0.44 &  9.38 &  3.40          &  10.1995  &    0.31  & 1948-2010 &   -1.4 &   0.7 &  \nodata & \nodata &  2.00 & 24.06 &  0.34\\ 
   (2062) &  Aten              & 0.97 &  0.18 & 18.93 &  1.10          &    40.77  &    0.26  & 1955-2012 &   -7.5 &   2.4 &  \nodata & \nodata &  3.17 & 19.94 &  0.65\\ 
   (1620) &  Geographos        & 1.25 &  0.34 & 13.34 &  2.56          &  5.22204  &  0.3258  & 1951-2012 &   -2.4 &   0.7 &   -2.5   &     0.6 &  3.85 & 18.15 &  0.48\\ 
   (2340) &  Hathor            & 0.84 &  0.45 &  5.85 &  0.30          &   \nodata &  \nodata & 1976-2012 &  -14.5 &   3.5 &  \nodata & \nodata &  4.11 & 15.32 &  0.31\\ 
   (1580) &  Betulia           & 2.20 &  0.49 & 52.11 &  5.80          &   6.1324  &    0.08  & 1950-2010 &   -1.4 &   2.0 &   -1.3   &     0.9 &  1.46 & 13.45 &  0.53\\ 
   (3361) &  Orpheus           & 1.21 &  0.32 &  2.69 &  0.30          &     3.58  &  \nodata & 1982-2009 &    5.7 &   2.5 &  \nodata & \nodata &  2.25 & 13.04 &  0.13\\ 
   (1566) &  Icarus            & 1.08 &  0.83 & 22.83 &  1.00          &    2.273  &    0.51  & 1949-2009 &   -3.2 &   2.0 &  \nodata & \nodata &  1.62 & 11.86 &  0.14\\ 
   (2063) &  Bacchus           & 1.08 &  0.35 &  9.43 &  1.35$\dagger$ &    14.90  &  \nodata & 1977-2007 &   -4.2 &   3.3 &  \nodata & \nodata &  1.26 & 10.58 &  0.42\\ 
    (887) &  Alinda            & 2.48 &  0.57 &  9.36 &  4.20          &    73.97  &    0.31  & 1918-2008 &    4.0 &   2.5 &  \nodata & \nodata &  1.59 &  9.42 &  1.12\\ 
   (1865) &  Cerberus          & 1.08 &  0.47 & 16.10 &  1.20          &    6.810  &    0.22  & 1971-2008 &   -5.1 &   2.7 &  \nodata & \nodata &  1.90 &  9.20 &  0.44\\ 
 (161989) &  Cacus             & 1.12 &  0.21 & 26.06 &  1.90          &   3.7538  &    0.09  & 1978-2010 &    2.6 &   2.3 &  \nodata & \nodata &  1.12 &  8.94 &  0.39\\ 
   (3757) &  1982 XB           & 1.83 &  0.45 &  3.87 &  0.50          &   9.0046  &    0.18  & 1982-2008 &  -13.7 &   6.8 &  \nodata & \nodata &  2.04 &  8.82 &  0.49\\ 
   (6037) &  1988 EG           & 1.27 &  0.50 &  3.50 &  0.65$\dagger$ &    2.760  &  \nodata & 1988-2007 &  -14.2 &   4.3 &  \nodata & \nodata &  3.34 &  8.51 &  0.64\\ 
   (2102) &  Tantalus          & 1.29 &  0.30 & 64.01 &  2.04$\dagger$ &    2.391  &  \nodata & 1975-2008 &   -3.8 &   3.6 &  \nodata & \nodata &  1.08 &  8.31 &  0.60\\ 
   (2100) &  Ra-Shalom         & 0.83 &  0.44 & 15.76 &  2.30          &   19.797  &    0.13  & 1975-2009 &   -4.8 &   2.2 &   -5.4   &     1.5 &  3.67 &  8.30 &  0.90\\ 
 (207945) &  1991 JW           & 1.04 &  0.12 &  8.71 &  0.52$\dagger$ &   \nodata &  \nodata & 1955-2009 &   -6.2 &   4.3 &  \nodata & \nodata &  1.42 &  8.00 &  0.26\\ 
  (67399) &  2000 PJ6          & 1.30 &  0.35 & 14.69 &  0.96$\dagger$ &   \nodata &  \nodata & 1951-2009 &  -19.4 &   7.2 &  \nodata & \nodata &  2.71 &  7.34 &  1.40\\ 
   (1036) &  Ganymed           & 2.66 &  0.53 & 26.70 & 31.66          &    10.31  &  0.2926  & 1924-2012 &   -6.6 &   1.5 &  \nodata & \nodata &  4.41 &  7.28 & 14.23\\ 
  (14402) &  1991 DB           & 1.72 &  0.40 & 11.42 &  0.60          &    2.266  &    0.14  & 1991-2009 &   -5.0 &   4.3 &  \nodata & \nodata &  1.19 &  7.05 &  0.22\\ 
   (3908) &  Nyx               & 1.93 &  0.46 &  2.18 &  1.00          &  4.42601  &    0.23  & 1980-2009 &    9.8 &   3.2 &   12.9   &     2.7 &  4.71 &  5.52 &  0.92\\ 
   (4660) &  Nereus            & 1.49 &  0.36 &  1.43 &  0.33          &     15.1  &    0.55  & 1981-2010 &    7.3 &   5.6 &   10.9   &     4.8 &  2.29 &  5.46 &  0.27\\ 
   (5660) &  1974 MA           & 1.79 &  0.76 & 38.06 &  2.57$\dagger$ &     17.5  &  \nodata & 1974-2005 &  -20.1 &  10.4 &  \nodata & \nodata &  1.92 &  5.46 &  2.68\\ 
  (10302) &  1989 ML           & 1.27 &  0.14 &  4.38 &  0.45$\dagger$ &      19.  &  \nodata & 1989-2006 &   35.3 &   7.1 &  \nodata & \nodata &  4.96 &  5.33 &  1.26\\ 
   (2061) &  Anza              & 2.26 &  0.54 &  3.77 &  2.60          &    11.50  &  \nodata & 1960-2012 &   10.7 &   9.0 &  \nodata & \nodata &  1.19 &  5.00 &  1.88\\ 
  (22753) &  1998 WT           & 1.22 &  0.57 &  3.20 &  1.02$\dagger$ &    10.24  &  \nodata & 1955-2009 &   -5.4 &   5.0 &   -6.1   &     4.9 &  1.26 &  4.95 &  0.41\\ 
   (3753) &  Cruithne          & 1.00 &  0.51 & 19.81 &  3.39$\dagger$ &     27.4  &  \nodata & 1973-2010 &  -11.2 &   5.3 &  \nodata & \nodata &  2.12 &  4.84 &  2.61\\ 
   (4769) &  Castalia          & 1.06 &  0.48 &  8.89 &  1.40          &    4.095  &  \nodata & 1989-2011 &  -17.2 &  11.7 &  \nodata & \nodata &  1.47 &  4.59 &  1.69\\ 
   (4947) &  Ninkasi           & 1.37 &  0.17 & 15.65 &  0.65$\dagger$ &   \nodata &  \nodata & 1978-2009 &   13.4 &  10.3 &  \nodata & \nodata &  1.30 &  4.23 &  0.69\\ 
 (241596) &  1998 XM2          & 1.80 &  0.34 & 27.10 &  1.41$\dagger$ &   \nodata &  \nodata & 1952-2011 &  -14.4 &  10.7 &  \nodata & \nodata &  1.35 &  4.23 &  1.53\\ 
   (4034) &  Vishnu            & 1.06 &  0.44 & 11.17 &  0.42          &   \nodata &    0.52  & 1986-2009 &  -13.8 &  12.1 &  \nodata & \nodata &  1.14 &  3.72 &  0.42\\ 
   (7336) &  Saunders          & 2.31 &  0.48 &  7.17 &  0.65$\dagger$ &    6.423  &  \nodata & 1982-2010 &   10.3 &   8.3 &  \nodata & \nodata &  1.25 &  3.50 &  0.47\\ 
   (2059) &  Baboquivari       & 2.64 &  0.53 & 11.04 &  2.46$\dagger$ &   \nodata &  \nodata & 1963-2009 &   17.7 &  12.8 &  \nodata & \nodata &  1.38 &  3.42 &  2.96\\ 
  (15745) &  1991 PM5          & 1.72 &  0.25 & 14.42 &  0.98$\dagger$ &   \nodata &  \nodata & 1982-2007 &  -13.2 &   9.0 &  \nodata & \nodata &  1.46 &  3.39 &  1.00\\ 
 (138911) &  2001 AE2          & 1.35 &  0.08 &  1.66 &  0.56$\dagger$ &   \nodata &  \nodata & 1984-2012 &  -22.9 &  11.2 &  \nodata & \nodata &  2.04 &  3.38 &  1.02\\ 
  (18109) &  2000 NG11         & 1.88 &  0.37 &  0.81 &  1.12$\dagger$ &   4.2534  &  \nodata & 1951-2005 &   12.0 &   9.6 &  \nodata & \nodata &  1.25 &  3.21 &  1.00\\ 
   (2202) &  Pele              & 2.29 &  0.51 &  8.74 &  1.07$\dagger$ &   \nodata &  \nodata & 1972-2008 &   29.5 &  21.2 &  \nodata & \nodata &  1.39 &  2.98 &  2.18\\ 
  (68950) &  2002 QF15         & 1.06 &  0.34 & 25.16 &  2.03$\dagger$ &      29.  &  \nodata & 1955-2008 &  -11.6 &   6.5 &  \nodata & \nodata &  1.80 &  2.96 &  1.78\\ 
   (4197) &  1982 TA           & 2.30 &  0.77 & 12.57 &  1.80          &   3.5380  &    0.37  & 1954-2010 &   30.9 &   9.2 &  \nodata & \nodata &  3.36 &  2.88 &  2.84\\ 
   (3362) &  Khufu             & 0.99 &  0.47 &  9.92 &  0.70          &   \nodata &    0.21  & 1984-2004 &  -20.4 &  13.2 &  \nodata & \nodata &  1.54 &  2.87 &  1.01\\ 
  (99935) &  2002 AV4          & 1.65 &  0.64 & 12.76 &  2.46$\dagger$ &   \nodata &  \nodata & 1955-2011 &   -9.8 &   8.0 &  \nodata & \nodata &  1.23 &  2.73 &  1.48\\ 
  (68350) &  2001 MK3          & 1.67 &  0.25 & 29.56 &  2.43$\dagger$ &     3.24  &  \nodata & 1955-2007 &  -14.4 &  11.9 &  \nodata & \nodata &  1.21 &  2.61 &  2.73\\ 
  (54690) &  2001 EB           & 1.63 &  0.26 & 35.36 &  1.18$\dagger$ &   \nodata &  \nodata & 1952-2009 &  -14.4 &  13.3 &  \nodata & \nodata &  1.08 &  2.56 &  1.31\\ 
   (4179) &  Toutatis          & 2.53 &  0.63 &  0.45 &  5.40          &     176.  &  \nodata & 1976-2011 &  -18.4 &   4.3 &   -5.0   &     0.6 &  8.33 &  2.44 &  1.68\\ 
   (1864) &  Daedalus          & 1.46 &  0.61 & 22.20 &  3.70          &    8.572  &  \nodata & 1971-2006 &  -16.9 &   8.2 &  \nodata & \nodata &  2.06 &  2.44 &  3.97\\ 
 (154330) &  2002 VX94         & 1.48 &  0.41 &  7.16 &  0.90$\dagger$ &   \nodata &  \nodata & 1986-2011 &   70.2 &  20.6 &  \nodata & \nodata &  3.42 &  2.41 &  4.64\\ 
   (7753) &  1988 XB           & 1.47 &  0.48 &  3.12 &  0.68$\dagger$ &   \nodata &  \nodata & 1988-2012 &   18.9 &  12.9 &  \nodata & \nodata &  1.46 &  2.39 &  0.90\\ 
  (10563) &  Izhdubar          & 1.01 &  0.27 & 63.46 &  1.48$\dagger$ &   \nodata &  \nodata & 1991-2010 &   32.3 &  14.5 &  \nodata & \nodata &  2.23 &  2.19 &  3.70\\ 
  (13651) &  1997 BR           & 1.34 &  0.31 & 17.25 &  1.07$\dagger$ &   33.644  &  \nodata & 1980-2011 &  -12.4 &  11.8 &  \nodata & \nodata &  1.06 &  2.18 &  1.02\\ 
  (12923) &  Zephyr            & 1.96 &  0.49 &  5.29 &  2.14$\dagger$ &    3.891  &  \nodata & 1955-2012 &  -26.5 &  12.1 &  \nodata & \nodata &  2.19 &  2.05 &  3.97\\ 
   (3554) &  Amun              & 0.97 &  0.28 & 23.36 &  2.48          &    2.530  &  0.1284  & 1986-2012 &    9.0 &   8.8 &  \nodata & \nodata &  1.03 &  2.05 &  1.73\\ 
  (88254) &  2001 FM129        & 1.18 &  0.63 &  1.52 &  1.19$\dagger$ &   \nodata &  \nodata & 1978-2008 &  -40.6 &  18.9 &  \nodata & \nodata &  2.15 &  2.05 &  3.01\\ 
\enddata

\tablenotetext{\!}{Orbital elements $a$, $e$, $i$ are from the MPCORB
  database.  Spin periods $P$ and geometric albedos $p_V$ are from the
  JPL Small-Body Database. Diameters $D$, when known, are from the
  same database, otherwise they are derived from the absolute
  magnitude with a $p_V=0.14$ assumption and marked with~$\dagger$.
  Objects are listed in decreasing order of Yarkvosky sensitivity
  $s_Y$.   Yarkosvky efficiencies $f_Y$ are estimated for a bulk
  density $\rho=1,200$ kg m$^{-3}$.}

\end{deluxetable}

\begin{table}[p]
\begin{center}
\caption{Statistical properties of observed Yarkovsky rates and efficiencies\label{tab-meanmedian}.}
 \begin{tabular}{lcccc}
\hline \hline
                  & \multicolumn{4}{c}{abs$(da/dt) \times 10^{-4}$ AU/Myr}\\
  Yarkovsky rate                  & mean & weighted mean & median & stdev \\ 
\hline
objects with $f_Y < 2\times10^{-5}$  & 7.6   &  4.4          &  5.6   &  6.4  \\
objects with $f_Y > 2\times10^{-5}$  & 27.0  & 18.5          & 20.1   & 18.7  \\
{\bf all objects}                    & 10.4  &  5.2          &  7.3   & 11.4  \\
\hline
                                  &\multicolumn{4}{c}{$f_Y \times 10^{-5}$}\\
  Yarkovsky efficiency            & mean & weighted mean & median & stdev \\ 
\hline
objects with $f_Y < 2\times10^{-5}$  & 0.67 &  0.53         & 0.50   & 0.51  \\
objects with $f_Y > 2\times10^{-5}$  & 4.50 &  7.47         & 3.01   & 3.38  \\
{\bf all objects}                    & 1.22 &  0.89         & 0.65   & 1.91  \\ 
\hline \hline
\end{tabular}
\end{center}
\end{table}

\end{document}